\documentclass[journal=jpclcd,manuscript=letter, layout=twocolumn]{achemso}

\usepackage[version=3]{mhchem} 
\usepackage{xr-hyper}
\usepackage{hyperref}
\usepackage{amsmath,amssymb,amsthm}
\usepackage[]{graphicx}
\usepackage{dblfloatfix}
\usepackage[]{abstract}

\let\oldmaketitle\maketitle
\let\maketitle\relax
    \makeatletter
    \newcommand*{\addFileDependency}[1]{
    \typeout{(#1)}
    \@addtofilelist{#1}
    \IfFileExists{#1}{}{\typeout{No file #1.}}
    }
    \makeatother

    \newcommand*{\myexternaldocument}[1]{%
    \externaldocument{#1}%
    \addFileDependency{#1.tex}%
    \addFileDependency{#1.aux}%
    }
    
    \myexternaldocument{auxFiles/supporting_JPCL}
\author{Enrico Trizio}
\affiliation[IIT]
{Atomistic Simulations, Istituto Italiano di Tecnologia, 16163, Genova, Italy}
\alsoaffiliation[UNIMIB]
{Department of Materials Science, Università di Milano-Bicocca, 20126, Milano, Italy}

\author{Michele Parrinello}
\email{michele.parrinello@iit.it}
\affiliation[IIT]
{Atomistic Simulations, Istituto Italiano di Tecnologia, 16163, Genova, Italy}

\title[From Enhanced Sampling to Reaction Profiles]
  {From Enhanced Sampling to Reaction Profiles}

\begin{document}

\twocolumn[
\oldmaketitle
\begin{abstract}
    The determination of efficient collective variables is crucial to the success of many enhanced sampling methods. As inspired by previous discrimination approaches, we first collect a set of data from the different metastable basins. The data are then projected with the help of a neural network into a low-dimensional manifold in which data from different basins are well discriminated. This is here guaranteed by imposing that the projected data follows a preassigned distribution. The collective variables thus obtained lead to an efficient sampling and often allow reducing the number of collective variables in a multi-basin scenario. We first check the validity of the method in two-state systems. We then move to multi-step chemical processes. In the latter case, at variance with previous approaches, one single collective variable suffices, leading not only to computational efficiency but to a very clear representation of the reaction free energy profile. 
\end{abstract}
]

        \begin{figure*}[b]
            \centering
            \includegraphics[width=0.85\textwidth]{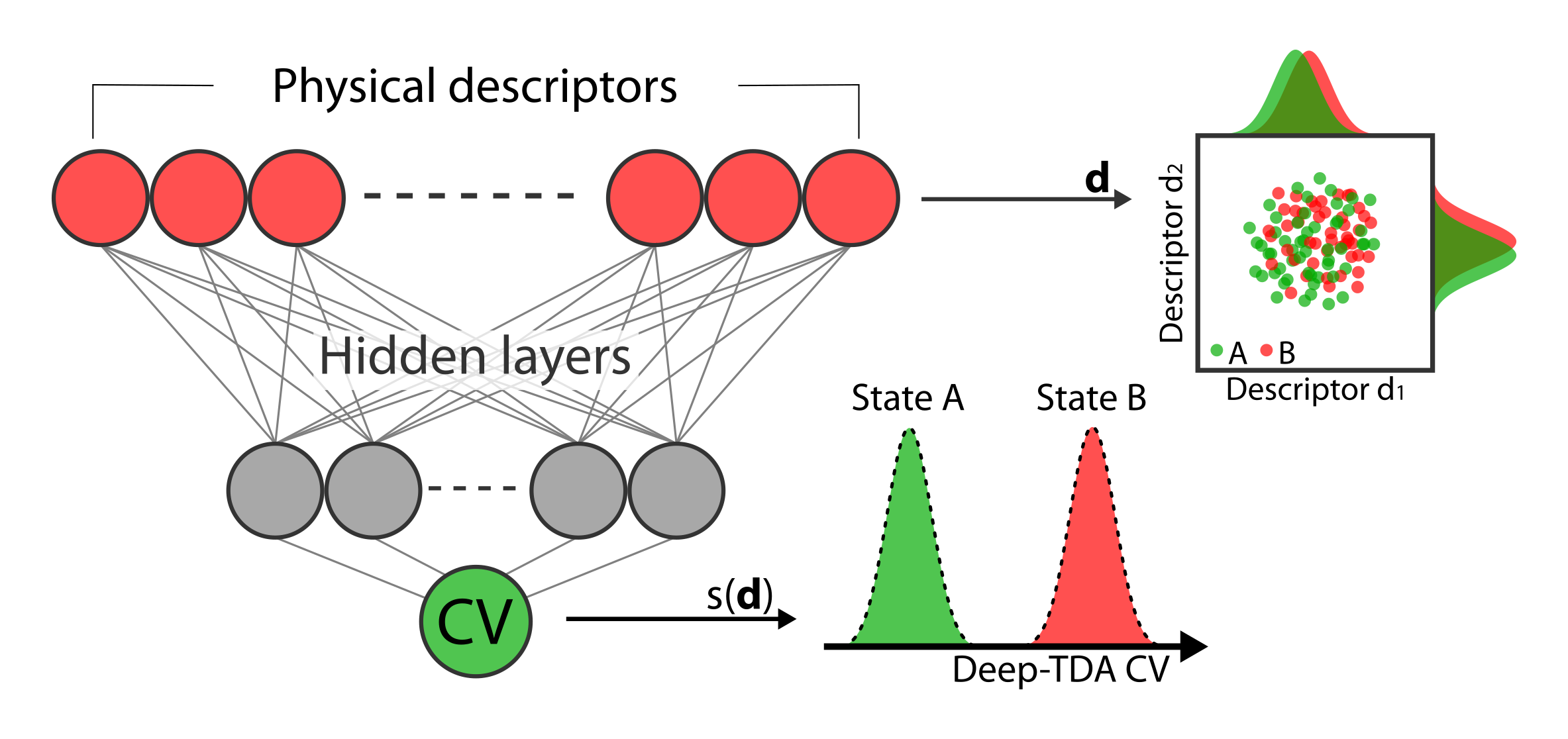}
            \caption{Schematic representation of the construction of the Deep-TDA CV. A set of physical descriptors \textbf{d} is fed as input of a feed-forward NN whose output layer directly gives the CVs. A series of nonlinear transformations through the NN hidden layers progressively compress the dimension using as objective function the distance in the projected space from a target distribution in which the states are well-discriminated. To further illustrate the process in the case of two states A and B, we have a sketch of the intricate data distribution with respect to some of the physical descriptors from the input set \textbf{d} and, below, the well-resolved distributions in the Deep-TDA CV $s(\textbf{d})$. }
            \label{fig:deepTDA_scheme}
        \end{figure*}

    Enhanced sampling methods are gaining increasing popularity because of their ability to alleviate the timescale problem, thus extending the scope of molecular dynamics (MD).
    Since the pioneering work of Torrie and Valleau\cite{torrie1977nonphysical}, many enhanced sampling methods have relied on the addition of a bias potential whose function is to promote transitions between different metastable states, eliminating kinetic bottlenecks \cite{laio2002escaping,barducci2008well,valsson2016enhancing}.
    The bias is taken to be a function of a small set of collective variables (CVs) $\{s\}$. 
    The CVs are in turn functions of the atomic coordinates $s=s(R)$ and encode the slow modes of the system. In addition, well-chosen CVs provide a coarse-grained but vivid description of a system's physics.
    The function of CV-based methods like umbrella sampling, metadynamics\cite{laio2002escaping,barducci2008well,valsson2016enhancing} or OPES\cite{invernizzi2020rethinking} is to enhance the fluctuations of the degrees of freedom coupled to $s(R)$, thus promoting rare event sampling.
    The CVs choice is thus of paramount importance for the success of this class of simulations and many methods have been proposed for their determination\cite{bussi2020using, tiwary2016spectral,ravindra2020automatic, m2017tica, wehmeyer2018time, chen2018molecular, peters2007extensions}.
    
    A typical scenario in which enhanced sampling methods are applied is the one in which there is a number of metastable states whose lifetime is so long that transitions between states are rare events unlikely to be observed in standard simulations. We shall characterize these states by a set of descriptors that are invariant with respect to the symmetry of the system. In such a way, states that differ only by translation, rotation or permutation of indistinguishable particles are not classified as different. Typical descriptors are distances, angles or coordination numbers. 
    In the multi-dimensional space of descriptors, configurations generated in unbiased runs in the different basins are well separated. When these data are mapped into the low-dimensional CV manifold, a necessary if not sufficient condition is that data belonging to different basins remain separated. Thus it has been suggested that discriminant analysis could be helpful in this context\cite{sultan2018automated,mendels2018collective,bonati2020data}.
    
    In the first applications of this approach, the time-honored  Linear Discriminant Analysis\cite{welling2005fisher} (LDA) method has been used to generate useful CVs \cite{mendels2018collective,piccini2018metadynamics,sultan2018automated}.
    However, often, when dealing with high-dimensional data, it is not possible to discriminate well the states using a linear scheme.
    Thus the method has been refined by using the hybrid Deep-LDA\cite{dorfer2015deep, bonati2020data} approach in which a strong non-linear feature is added to the standard-LDA scheme with the help of a Neural Network (NN).
    This method has proven highly effective in applications as diverse as chemical reactions\cite{bonati2020data}, crystallization\cite{karmakar2021colvarCryst} and ligand binding\cite{rizzi2021role}. This empirical evidence is reassuring since there is no guarantee a priori that in general a CV based on a discrimination criterion only is effective, since it does not explicitly encode information on the transition state. Furthermore, this approach is rather attractive since there is in theory no limit to the number of descriptors that can be used and the CV construction procedure is semi-automatic.
   
    Here, we first modify the Deep-LDA procedure\cite{bonati2020data,dorfer2015deep} by skipping the linear step altogether and express the CV directly as the NN output layer. When the objective function used for the NN optimization is of the Fisher type (Deep Discriminant Analysis, Deep-DA, see SI), we find results very close to those of Ref.\cite{bonati2020data,mendels2018collective,piccini2018metadynamics,rizzi2021role}. However, the loss function can be designed so as the discriminating variable distribution matches a preassigned target in which the states are well discriminated. We shall refer to this method as Deep Targeted Discriminant Analysis (Deep-TDA). 
    
    The advantages of Deep-TDA become clearer when it is applied to an important class of systems that have $N_S$ states. In such a case, the use of Deep-LDA forces one to use a CV that has $(N_S - 1)$ components. This leads to an increase in the computational cost that grows exponentially with $N_S$. 
    However, this is not necessary in the many cases in which a reaction proceeds through a well-defined succession of intermediate steps. 
    Under these circumstances, one can design the Deep-TDA target distribution able to discriminate the reaction steps only with one single CV. This not only reduces the computational impact of a multi-state calculation but leads to more easily interpretable results that can be represented as a reaction profile of the type familiar to quantum chemists.

    \paragraph{Method - Two-state scenario}
    As for Deep-LDA, given two states A and B, characterized by a set of descriptors \textbf{d}, we want to construct a CV $s$ by finding a one-dimensional projection along which the two states are well-discriminated.
    In the SI, we provide the details of a reformulation of the Deep-LDA procedure in which we skip the linear step and express $s$ directly as the output of a feed-forward NN optimized with a loss function of Fisher type (Deep-DA) as in Ref.\cite{bonati2020data,mendels2018collective,piccini2018metadynamics}. We also show that the performances of this non-linear variant of LDA are similar to those of standard Deep-LDA, as to be expected.
    
    Having satisfactorily shown that the linear step of Deep-LDA can be skipped, we reformulate the optimization criterion of the CV $s$ to make it more direct and flexible. 
    To do so, we train the NN so as the distribution of the metastable states along the CV follows a preassigned bimodal distribution (Deep-TDA, see Fig.\ref{fig:deepTDA_scheme}). 
    The target distribution to be used in this framework can be in principle of any type, as long as it guarantees proper discrimination between the states. A natural choice is to use as target two Gaussian distributions of preassigned positions and widths, such that the A configurations are distributed according to one of the two Gaussians and B configurations according to the other.
    The ability of a target distribution to discriminate between the two states can be measured with a Fisher-like ratio
        \begin{equation}
                F = \frac{(\mu_A - \mu_B)^2}{\sigma_A^2 + \sigma_B^2}
                \label{eq:fisher}
        \end{equation}
    where $\mu_A$ and $\mu_B$ are the average values for the two states in the CV space and $\sigma_A^2$ and $\sigma_B^2$ their variances.  
    However, as $F$ rapidly grows to very large values when the separation between the states is increased, we prefer to use $\Delta = \sqrt{F}$ as a parameter to characterize our target function. Because in the end we want to use $s$ as a CV and not only as a discriminator, the choice of such parameter requires some experimentation.
    Indeed, to be effective, $s$ needs to assume properly interpolating values when dealing with transition states. In the practice, this implies that $\Delta$ can be neither too small nor too large.
    If $\Delta$ is too small, the CV is not able to discriminate correctly between states. On the other hand, if $\Delta$ is too large, $s$ cannot describe well the transition state. As a rule of thumb, we found that values in the range $25<\Delta<50$ are appropriate choices. 
    
    As far as the objective function is concerned, we could have enforced the target distributions using Kulback-Leibler divergences. However, in the case of Gaussians, this would have been an overkill and thus we simply impose that the two distributions have preassigned positions and widths.
    Thus each state $k$ contributes to the loss function two terms, one that enforces its center $L^\mu_k$ and the other its width $L_k^\sigma$
        \begin{equation}
            L_k^\mu = (\mu_k - \mu_k^{tg})^2 \qquad\quad
            L_k^\sigma = \left(\sigma_k - \sigma_k^{tg}\right)^2
            \label{eq:loss_contributes}
        \end{equation}
    The whole loss function is then given by the linear combination of all the contributions.
        \begin{equation}
            L = \sum_{k}\alpha L_k^\mu + \beta L_k^\sigma \qquad k = A,B
            \label{eq:loss}
        \end{equation}
    in which the hyperparameters $\alpha$ and $\beta$ are chosen such that the two terms are scaled to roughly the same order of magnitude at the earlier stages of the optimization.
    Conveniently the loss function in Eq.\ref{eq:loss} tends to zero as we approach the target distribution, thus convergence is simple to monitor.
  
 \begin{figure*}[b!]
    \begin{minipage}{0.49\linewidth}
        \includegraphics[width=\linewidth]{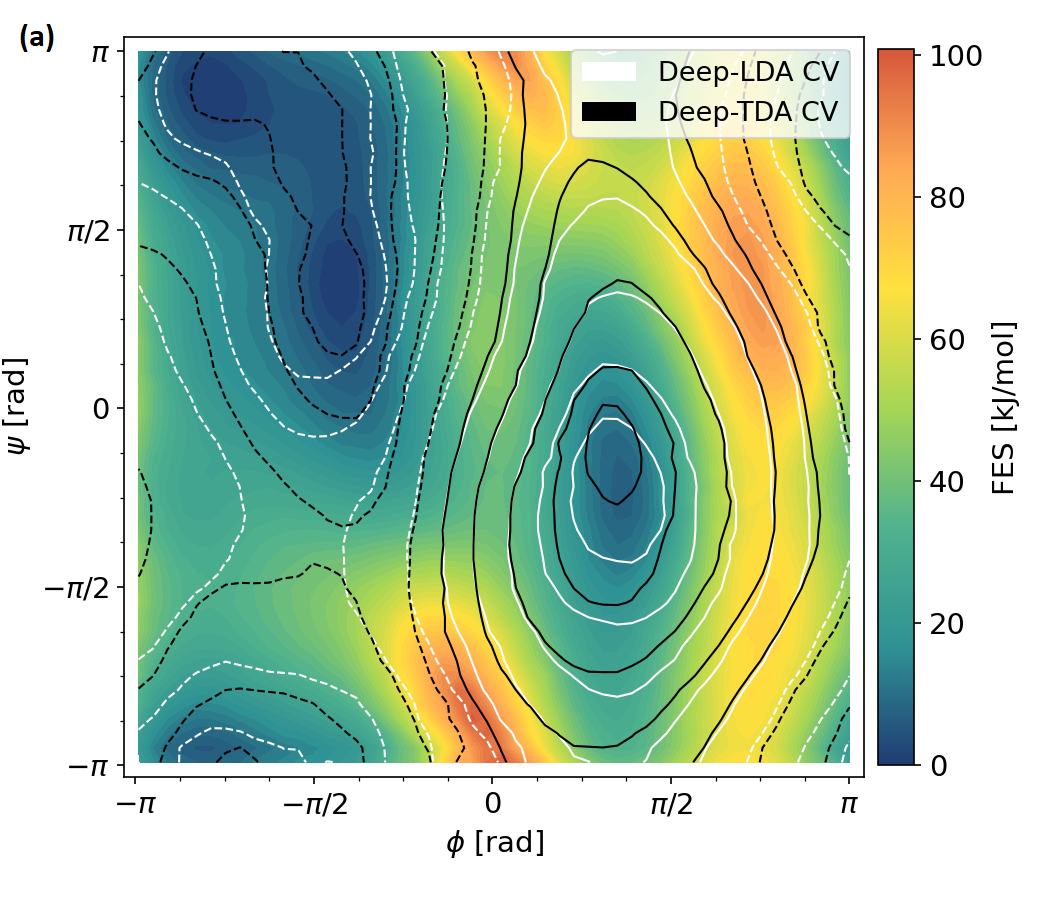}
    \end{minipage}
    \hspace*{\fill}
    \begin{minipage}{0.49\linewidth}
    \includegraphics[width=\linewidth]{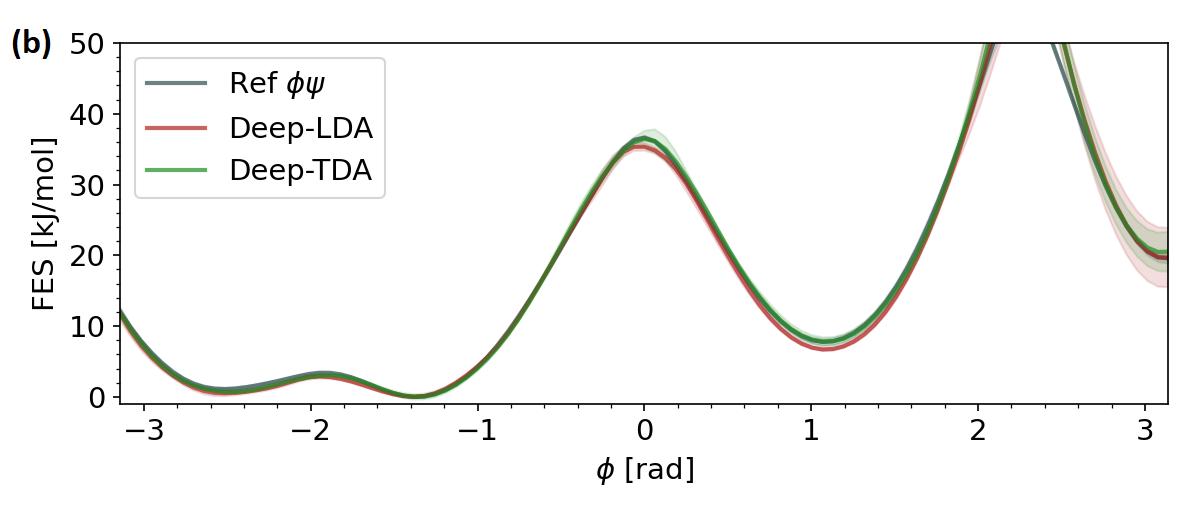}
    \includegraphics[width=\linewidth]{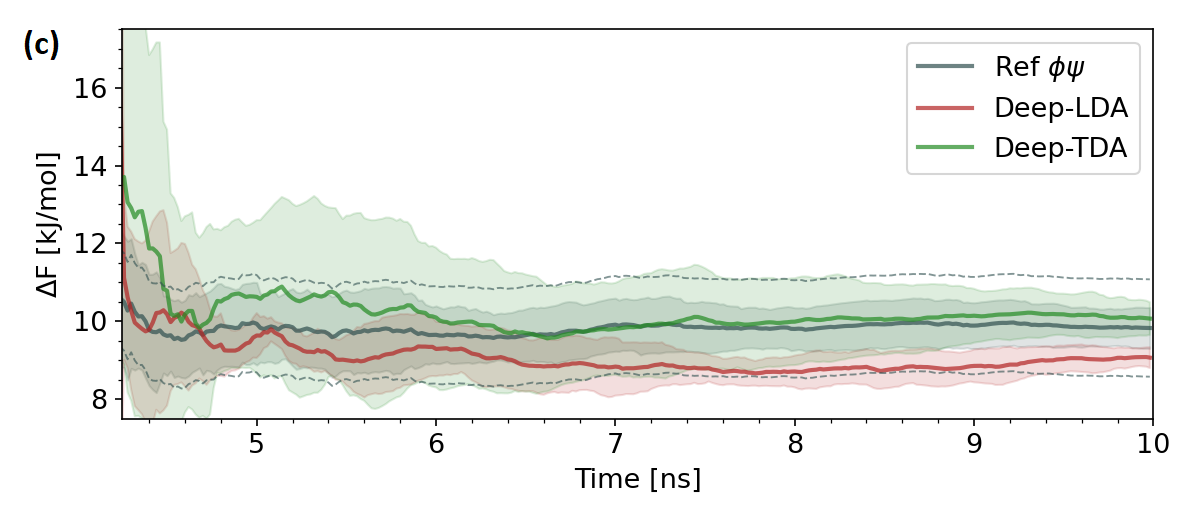}
    \end{minipage}
    \caption{Results of OPES enhanced simulations of the folding of alanine dipeptide. \textbf{(a)} Comparison of the isolines of Deep-LDA (white) and Deep-TDA (black) CVs on top of the energetic landscape in Ramachandran angles $\phi\psi$ plane. Solid lines are for positive CV values, dashed lines for negative ones.
    \textbf{(b-c)} Comparison of energy estimates from OPES simulations using Deep-LDA, Deep-TDA or a reference $\phi\psi$ as CVs. The results are averaged on five independent simulations for each CV to get the mean solid line and the standard deviation error bars. \textbf{(b)} FES profile estimates along the $\phi$ Ramachandran angle. \textbf{(c)} $\Delta$F between the metastable basins estimates, obtained as functions of the simulation time. The dashed lines give the $\pm0.5K_bT$ range on the reference curve.}
    \label{fig:alanine}
\end{figure*}

    \paragraph{Method - Multi-state scenario}
    
    The extension to the multi-state case is straightforward in the Deep-TDA framework. We recall that in a system with $N_S$ states in the general case one needs to define $(N_S-1)$ CVs. We build the CVs by imposing a target that is a linear superposition of $N_S$ multivariate Gaussians with diagonal covariances. Each Gaussian is then defined by $N_\rho = (N_S-1)$ CV positions and covariances, thus leading to the following loss function 
             \begin{equation}
                L = \sum_{k}^{N_S}\sum_{\rho}^{N_\rho}\alpha L_{k,\rho}^\mu + \beta L_{k,\rho}^\sigma 
                \label{eq:loss_multi}
            \end{equation}
    where $\rho$ are the components of the CVs space.
    The location of the different Gaussians is arbitrary, but as before attention has to be paid to the relative distances and widths.
 
    As anticipated in the introduction, there are circumstances in which using Deep-TDA one can reduce the number of CVs. This reduction is possible if the reaction involves a series of steps that one can sequentially align. 
    In these situations, the topology of the problem is actually linear and the number of CVs can be reduced to one with clear computational advantages. Of course, this reduction is only possible because we are using information on the dynamics of the system.

\paragraph{Results}

For didactical purposes, we report first the application of Deep-TDA to the study of alanine dipeptide in vacuum. This simple but instructive example will confirm that, in the two-state case, Deep-TDA works just as well as Deep-LDA, as to be expected. This is also confirmed by a more challenging application to a non-trivial host-guest problem that is reported in the SI. After having dealt with two-state test cases, we apply Deep-TDA to the multi-state case of the hydrobromination of propene and of a double intramolecular proton transfer reaction, where the advantage of using Deep-TDA becomes clearly apparent. The technical details of the calculation, as well as some more in-depth comparisons of Deep-TDA and Deep-LDA results, are reported in the SI.

    \paragraph{Alanine Dipeptide} 
        As it is well known, alanine dipeptide at room temperature is a two-state system. In this small molecule, the two Ramachandran angles $\phi$ and $\psi$ are known to be good CVs and we could have used these angles as descriptors. However, following Ref.\cite{bonati2020data}, we tested the robustness of the method by using as descriptors the 45 distances between the heavy atoms.  
        
        The isolines of Deep-TDA CV are reported in panel \textbf{(a)} of Fig.\ref{fig:alanine} and are equivalent to those of Deep-LDA and follow the underlying energetic landscape in the Ramachandran plot even far from the training basins.
        Using Deep-TDA CV and OPES effectively encourages many transitions between the two metastable basins. The performances are comparable to the nearly ideal set of Ramachandran angles as confirmed by the consistency of the estimates of the FES along the $\phi$ torsion angle and of the free energy difference between the basins, as shown in panels \textbf{(b-c)} of Fig.\ref{fig:alanine}.

\begin{figure} [h!]
                \centering                \includegraphics[width=0.9\linewidth]{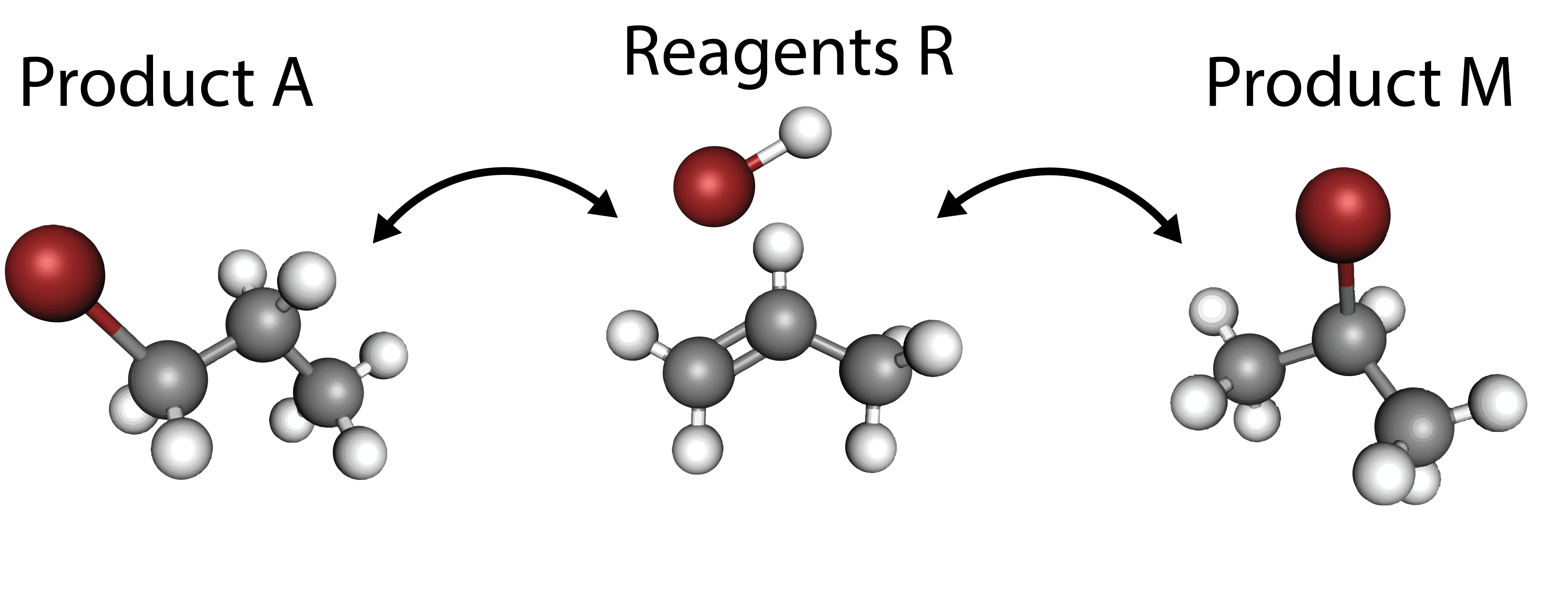}
                \caption{Reaction scheme for the hydrobromination of propene. The same reagents R can lead to two possible products, anti-Markovnikov A and Markovnikov M, depending on the addition position of the halide.}
                \label{fig:hbro_overview}
            \end{figure}

    \paragraph{Hydrobromination of propene}
        Having assessed that Deep-TDA is a good substitute for Deep-LDA, we now turn to discuss cases in which Deep-TDA can give a competitive advantage.
        One of them is the case of the hydrobromination of propene. Starting from the same reagents (R), this chemical reaction can lead to two products, identified as Markovnikov (M) and Anti-Markovnikov (A) (see Fig.\ref{fig:hbro_overview}).
                 
        This is clearly a multi-state scenario and we can tackle this problem in two ways. 
        One in which, following the prescription of LDA, we determine two CVs. The second takes advantage of the fact that products A and M interconvert with a very low probability. Thus one is in a scenario in which one can map the problem into a linear sequence of reactions $A \leftrightharpoons R \leftrightharpoons M$ and reduce the number of CVs from two to one, as anticipated earlier.
        
        We compare here the two possibilities, starting from the standard two CVs approach. In this case, we have chosen a very simple target that is a sum of multivariate Gaussian functions with diagonal covariances placed at the vertices of an equilateral triangle. The positions and widths of the Gaussians were chosen so as to satisfy the criteria described earlier.
         \begin{figure} [b!]
                \centering \includegraphics[width=\linewidth]{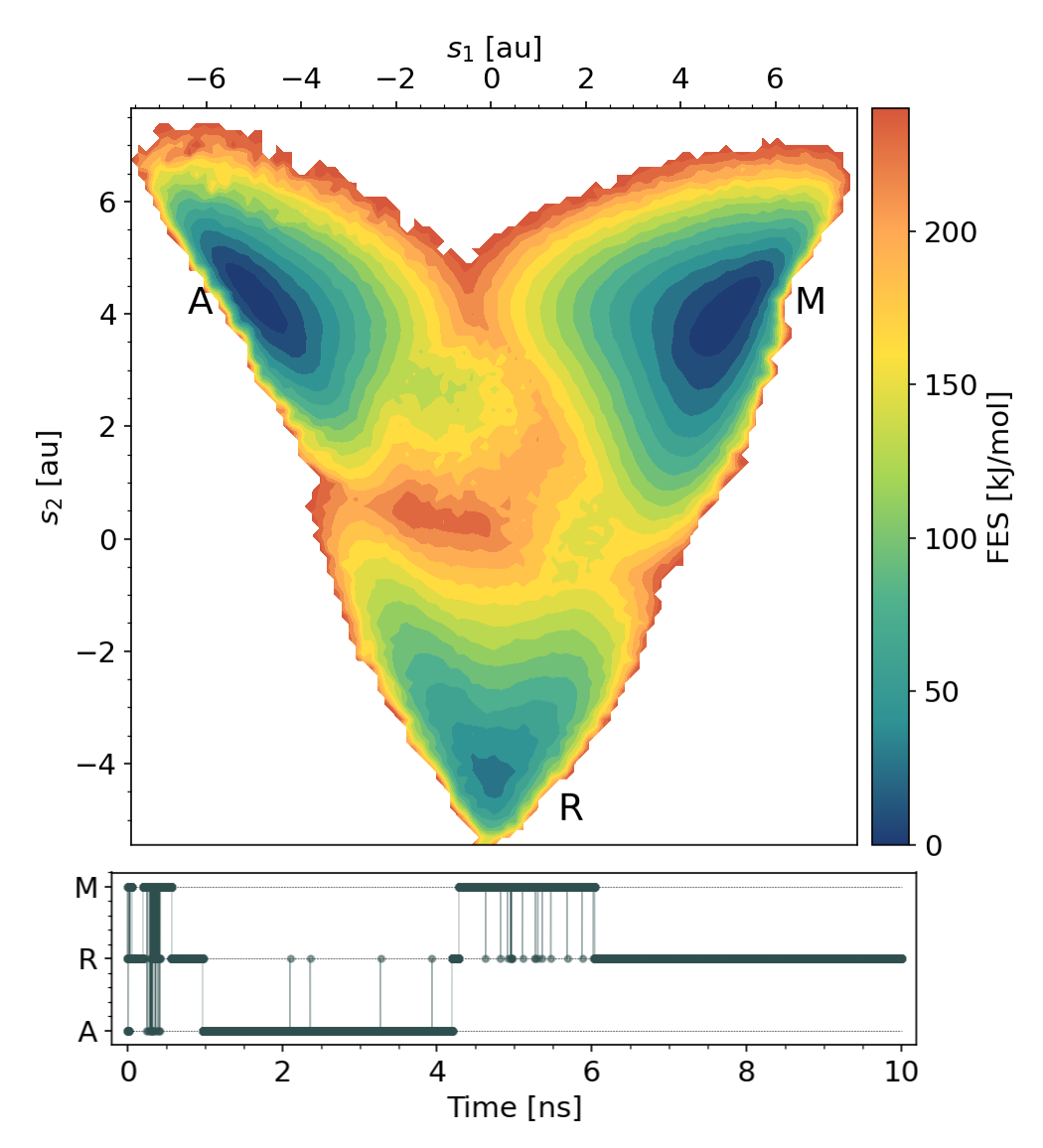}
                \caption{\textbf{Upper panel}: FES estimate for the hydrobromination of propene in the plane of the Deep-TDA CVs $s_1$ and $s_2$. The basins are labeled as anti-Markovnikov A, reagents R and Markovnikov M. In the SI, an error analysis can be found. 
                \textbf{Lower panel}: State occupation during a single 10ns OPES run with the two-dimensional Deep-TDA CV. As indicator functions, we use, for each metastable state, its descriptors density, modeled as a Gaussian mixture\cite{debnath2020gaussian}. The density was trained on the same unbiased data and uses the same descriptors employed in the generation of the Deep-TDA CV.} 
                \label{fig:hbro_2D}
            \end{figure}
        
        As in Ref.\cite{rizzi2019blind}, we used as descriptors the contact functions
        \begin{equation}
            c_{ij}(r) = \frac{1 - \left(\frac{r}{\sigma_{ij}}\right)^n}{1 - \left(\frac{r}{\sigma_{ij}}\right)^m}
            \label{eq:contact}
        \end{equation}
        where $r$ are the pairwise atomic distances and $\sigma_{ij}$ is the typical bond length between the involved species $i$ and $j$.
        In order to focus the sampling on the relevant reaction, we break the permutational symmetry and allow only the H in the initial HBr molecule to react\cite{debnath2020gaussian} (see SI).
        
        The two-dimensional free energy surface is shown in the upper panel of Fig.\ref{fig:hbro_2D} and reflects the expected free energy order of the different states. However, the non-linearity of the CVs distorts the FES and makes it difficult to identify the transition paths. In fact, an analysis of the OPES trajectories showed that the system never made a direct transition between $A$ and $M$. The only allowed transitions were $A\leftrightharpoons R\leftrightharpoons M$ (see the lower panel in Fig.\ref{fig:hbro_2D}).
        
        This suggested to use a one-dimensional CV, thus we designed a one-dimensional target that at the center has the reagents R and at the sides the products A and M and used our Deep-TDA method. 
        When used in  OPES, the CV was able to drive transitions from reactants to products without attempting any $A\leftrightharpoons M$ direct  transition (see SI).
            \begin{figure}[b!]
                \centering               
                \includegraphics[width=\linewidth]{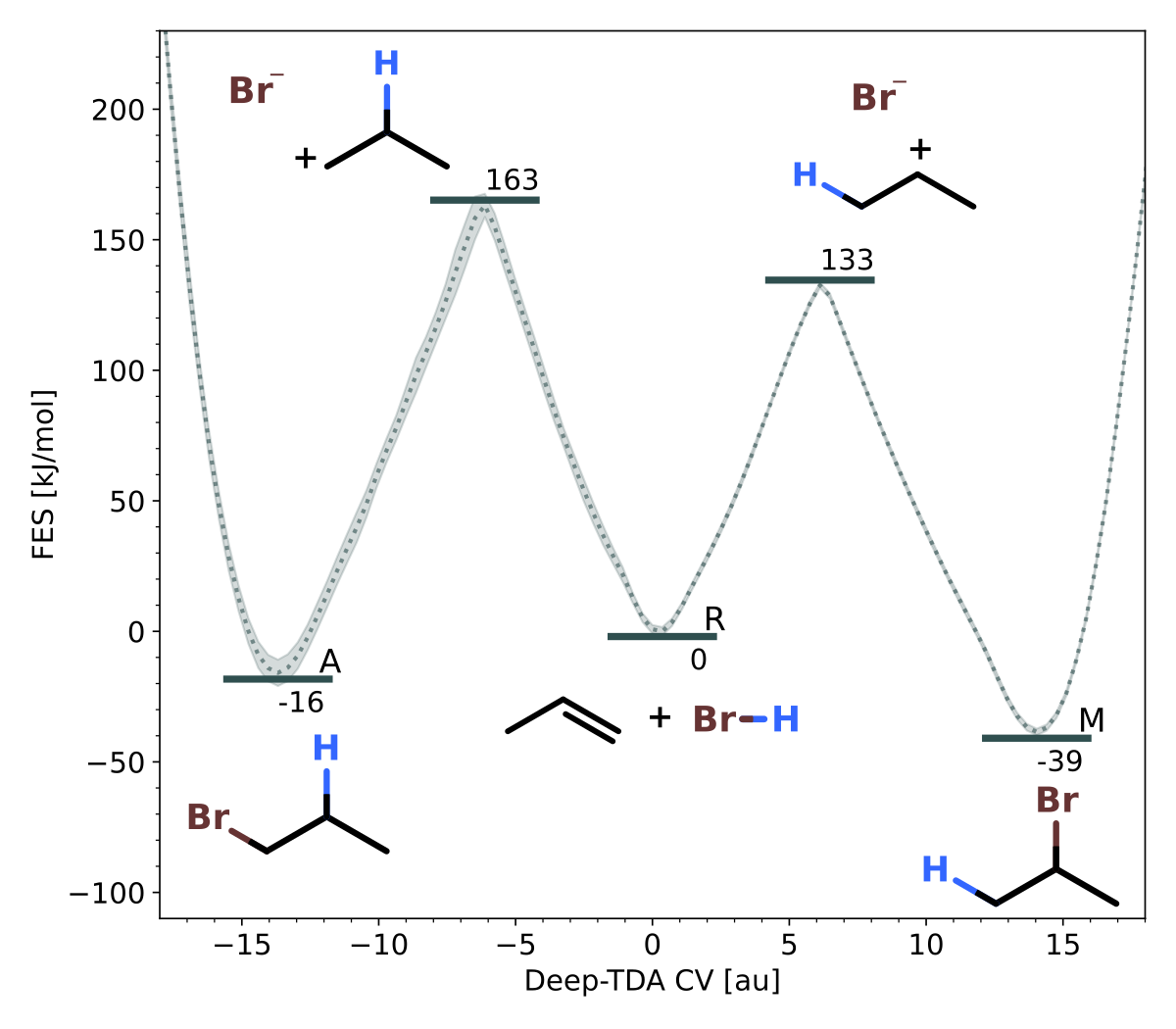}
                \caption{ 
                FES estimate profile for the multi-state hydrobromination of propene reaction projected along the Deep-TDA CV, with the indication of the different metastable and transition states. The dotted line gives the average FES profile with the related error. The free energy of each state is expressed in kJ/mol.}
                \label{fig:hbro_fes_1D}
            \end{figure}
            
        The resulting one-dimensional FES in Fig.\ref{fig:hbro_fes_1D} is more easily readable than the two-dimensional one in Fig.\ref{fig:hbro_2D}, with the different states and transition states clearly marked as done in the standard representation of chemical processes.
        This representation also illuminates the fact that the selectivity towards the Markovnikov product is due to the kinetics of the process rather than to its thermodynamics.

    \paragraph{Double proton transfer in diamino-benzoquinone}
        
    \begin{figure}[b!]
                \centering 
                \includegraphics[width=\linewidth]{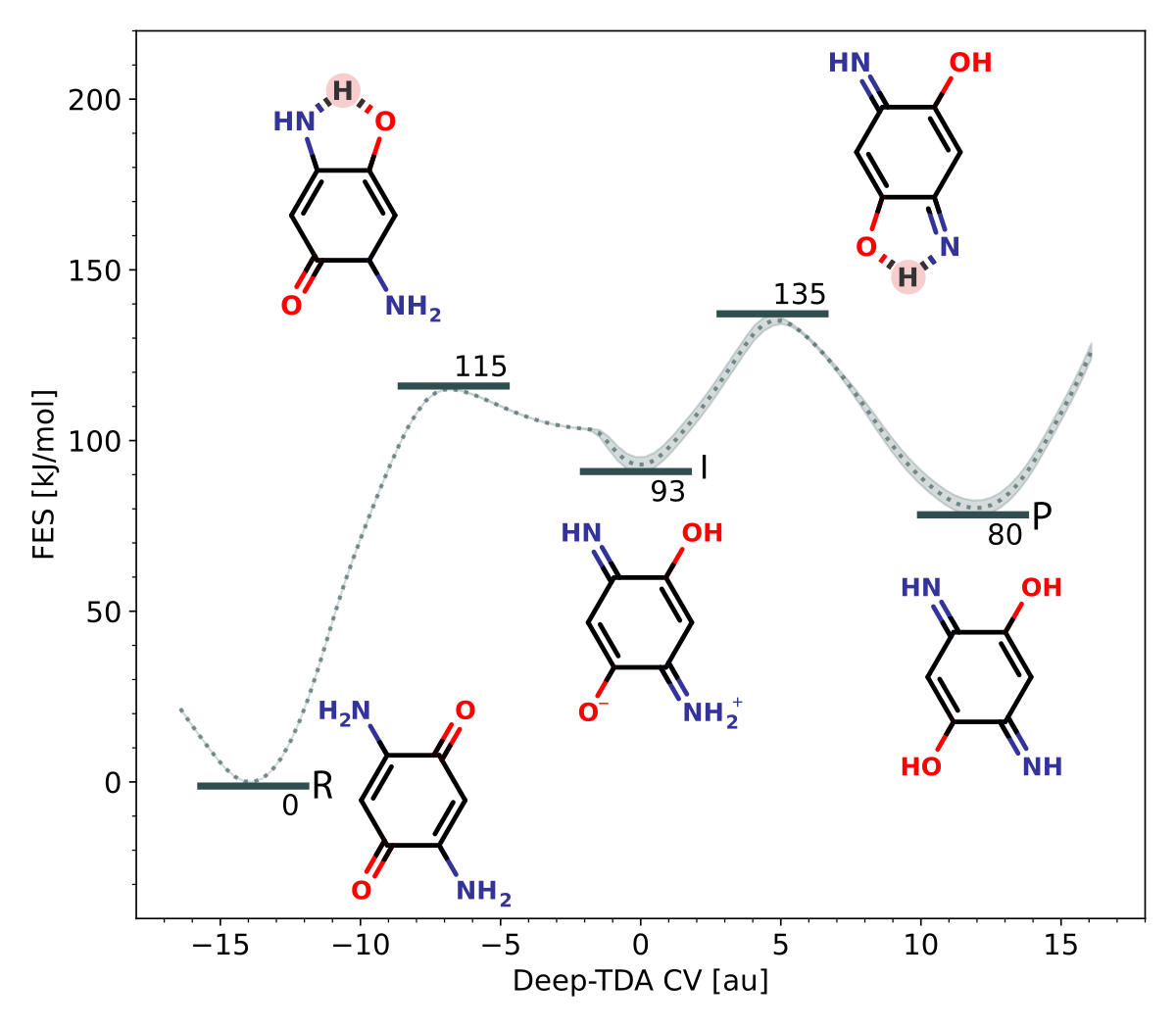}
                \caption{FES for the double intramolecular proton transfer reaction in 2,5-diamino-1,4-benzoquinone projected along the Deep-TDA CV. The metastable states are, from left to right: keto form (\textbf{R}), intermediate (\textbf{I}) and enol form (\textbf{P}). The dotted line gives the average FES profile with the related error. The free energy of each state is expressed in kJ/mol.}
                \label{fig:benzo_fes_1D}
            \end{figure}  
                
        The one-dimensional approach can also be applied 
        to chemical reactions in which a number of reactive steps take place in a well-defined order. One such case is the double proton transfer of 2,5-diamino-1,4-benzoquinone. For this reaction, a two-step mechanism has been proposed\cite{rumpel1989nmr, fatollahpour2017dft} with a stable intermediate (I) between the reagent (R) and product (P), which are respectively the enol and keto stable forms of the compound (see Fig.\ref{fig:benzo_fes_1D}).
        To determine the one-dimensional Deep-TDA CV, we have thus designed a target with the correct ordering of the states so as to account for transitions of the type
        $R\leftrightharpoons I\leftrightharpoons P$.
        As descriptors, we have used the heavy atom coordination numbers to preserve all the symmetries of the system (see SI for details).

        When used in biased OPES simulations, also here the Deep-TDA CV was able to promote efficiently the different reaction steps.  Also in this case the resulting one-dimensional FES (Fig.\ref{fig:benzo_fes_1D}) clearly shows the three metastable states and gives a neat representation of the reaction profile with free energies compatible with those obtained from static calculations\cite{fatollahpour2017dft}.
        In biased runs, the system was also able to explore less likely rotational isomers, different from the dominant ones in the training set (see SI), with a slight effect on the shape of the I and P basins. 
    
\paragraph{Conclusions}
    In this paper, we have shown that our Deep-TDA method has several interesting properties and has a flexibility that Deep-LDA does not have.
    In the general case, it does perform just as well but offers an efficient and physically transparent alternative in the multi-state scenario.
    
    We wish to remark that the model aims not at finding the best CV for a given system but rather to provide a reliable one starting from limited information and that such CV is suitable to be used in combination with any CV-based method.
    
    Considering all the reasons above, we think that this model could have some use in the study of a broad range of rare events, especially when multiple states are involved.

\paragraph{Acknowledgements}

The authors want to thank Luigi Bonati, Davide Mandelli and Umberto Raucci for all the useful discussions and suggestions; Andrea Rizzi for thoughtfully reading the paper; Jayashrita Debnath, Michele Invernizzi and Valerio Rizzi for their help in finding the best simulation conditions.

\paragraph{Supplementary information}

The complete computational details are given in the SI.
All the reported enhanced sampling simulations have been carried out using the open-source plug-in PLUMED-2.7\cite{bonomi2009plumed,tribello2014plumed,bonomi2019promoting}. This has been patched with the Gromacs-2019.6\cite{abraham2015gromacs} MD engine for the simulations for alanine dipeptide and with CP2K-7.1\cite{kuhne2020cp2k} for the hydrobromination and proton transfer simulations. 
The training of the CVs has been done relying on the Pytorch\cite{paszke2017automatic} library.

\bibliography{bibliography.bib}

\providecommand{\latin}[1]{#1}
\makeatletter
\providecommand{\doi}
  {\begingroup\let\do\@makeother\dospecials
  \catcode`\{=1 \catcode`\}=2 \doi@aux}
\providecommand{\doi@aux}[1]{\endgroup\texttt{#1}}
\makeatother
\providecommand*\mcitethebibliography{\thebibliography}
\csname @ifundefined\endcsname{endmcitethebibliography}
  {\let\endmcitethebibliography\endthebibliography}{}
\begin{mcitethebibliography}{31}
\providecommand*\natexlab[1]{#1}
\providecommand*\mciteSetBstSublistMode[1]{}
\providecommand*\mciteSetBstMaxWidthForm[2]{}
\providecommand*\mciteBstWouldAddEndPuncttrue
  {\def\EndOfBibitem{\unskip.}}
\providecommand*\mciteBstWouldAddEndPunctfalse
  {\let\EndOfBibitem\relax}
\providecommand*\mciteSetBstMidEndSepPunct[3]{}
\providecommand*\mciteSetBstSublistLabelBeginEnd[3]{}
\providecommand*\EndOfBibitem{}
\mciteSetBstSublistMode{f}
\mciteSetBstMaxWidthForm{subitem}{(\alph{mcitesubitemcount})}
\mciteSetBstSublistLabelBeginEnd
  {\mcitemaxwidthsubitemform\space}
  {\relax}
  {\relax}

\bibitem[Torrie and Valleau(1977)Torrie, and Valleau]{torrie1977nonphysical}
Torrie,~G.~M.; Valleau,~J.~P. Nonphysical sampling distributions in {M}onte
  {C}arlo free-energy estimation: Umbrella sampling. \emph{Journal of
  Computational Physics} \textbf{1977}, \emph{23}, 187--199\relax
\mciteBstWouldAddEndPuncttrue
\mciteSetBstMidEndSepPunct{\mcitedefaultmidpunct}
{\mcitedefaultendpunct}{\mcitedefaultseppunct}\relax
\EndOfBibitem
\bibitem[Laio and Parrinello(2002)Laio, and Parrinello]{laio2002escaping}
Laio,~A.; Parrinello,~M. Escaping free-energy minima. \emph{Proceedings of the
  National Academy of Sciences} \textbf{2002}, \emph{99}, 12562--12566\relax
\mciteBstWouldAddEndPuncttrue
\mciteSetBstMidEndSepPunct{\mcitedefaultmidpunct}
{\mcitedefaultendpunct}{\mcitedefaultseppunct}\relax
\EndOfBibitem
\bibitem[Barducci \latin{et~al.}(2008)Barducci, Bussi, and
  Parrinello]{barducci2008well}
Barducci,~A.; Bussi,~G.; Parrinello,~M. Well-tempered metadynamics: a smoothly
  converging and tunable free-energy method. \emph{Physical review letters}
  \textbf{2008}, \emph{100}, 020603\relax
\mciteBstWouldAddEndPuncttrue
\mciteSetBstMidEndSepPunct{\mcitedefaultmidpunct}
{\mcitedefaultendpunct}{\mcitedefaultseppunct}\relax
\EndOfBibitem
\bibitem[Valsson \latin{et~al.}(2016)Valsson, Tiwary, and
  Parrinello]{valsson2016enhancing}
Valsson,~O.; Tiwary,~P.; Parrinello,~M. Enhancing important fluctuations: Rare
  events and metadynamics from a conceptual viewpoint. \emph{Annual review of
  physical chemistry} \textbf{2016}, \emph{67}, 159--184\relax
\mciteBstWouldAddEndPuncttrue
\mciteSetBstMidEndSepPunct{\mcitedefaultmidpunct}
{\mcitedefaultendpunct}{\mcitedefaultseppunct}\relax
\EndOfBibitem
\bibitem[Invernizzi and Parrinello(2020)Invernizzi, and
  Parrinello]{invernizzi2020rethinking}
Invernizzi,~M.; Parrinello,~M. Rethinking metadynamics: from bias potentials to
  probability distributions. \emph{The journal of physical chemistry letters}
  \textbf{2020}, \emph{11}, 2731--2736\relax
\mciteBstWouldAddEndPuncttrue
\mciteSetBstMidEndSepPunct{\mcitedefaultmidpunct}
{\mcitedefaultendpunct}{\mcitedefaultseppunct}\relax
\EndOfBibitem
\bibitem[Bussi and Laio(2020)Bussi, and Laio]{bussi2020using}
Bussi,~G.; Laio,~A. Using metadynamics to explore complex free-energy
  landscapes. \emph{Nature Reviews Physics} \textbf{2020}, \emph{2},
  200--212\relax
\mciteBstWouldAddEndPuncttrue
\mciteSetBstMidEndSepPunct{\mcitedefaultmidpunct}
{\mcitedefaultendpunct}{\mcitedefaultseppunct}\relax
\EndOfBibitem
\bibitem[Tiwary and Berne(2016)Tiwary, and Berne]{tiwary2016spectral}
Tiwary,~P.; Berne,~B. Spectral gap optimization of order parameters for
  sampling complex molecular systems. \emph{Proceedings of the National Academy
  of Sciences} \textbf{2016}, \emph{113}, 2839--2844\relax
\mciteBstWouldAddEndPuncttrue
\mciteSetBstMidEndSepPunct{\mcitedefaultmidpunct}
{\mcitedefaultendpunct}{\mcitedefaultseppunct}\relax
\EndOfBibitem
\bibitem[Ravindra \latin{et~al.}(2020)Ravindra, Smith, and
  Tiwary]{ravindra2020automatic}
Ravindra,~P.; Smith,~Z.; Tiwary,~P. Automatic mutual information noise omission
  (AMINO): generating order parameters for molecular systems. \emph{Molecular
  Systems Design \& Engineering} \textbf{2020}, \emph{5}, 339--348\relax
\mciteBstWouldAddEndPuncttrue
\mciteSetBstMidEndSepPunct{\mcitedefaultmidpunct}
{\mcitedefaultendpunct}{\mcitedefaultseppunct}\relax
\EndOfBibitem
\bibitem[M.~Sultan and Pande(2017)M.~Sultan, and Pande]{m2017tica}
M.~Sultan,~M.; Pande,~V.~S. TICA-metadynamics: accelerating metadynamics by
  using kinetically selected collective variables. \emph{Journal of chemical
  theory and computation} \textbf{2017}, \emph{13}, 2440--2447\relax
\mciteBstWouldAddEndPuncttrue
\mciteSetBstMidEndSepPunct{\mcitedefaultmidpunct}
{\mcitedefaultendpunct}{\mcitedefaultseppunct}\relax
\EndOfBibitem
\bibitem[Wehmeyer and No{\'e}(2018)Wehmeyer, and No{\'e}]{wehmeyer2018time}
Wehmeyer,~C.; No{\'e},~F. Time-lagged autoencoders: Deep learning of slow
  collective variables for molecular kinetics. \emph{The Journal of chemical
  physics} \textbf{2018}, \emph{148}, 241703\relax
\mciteBstWouldAddEndPuncttrue
\mciteSetBstMidEndSepPunct{\mcitedefaultmidpunct}
{\mcitedefaultendpunct}{\mcitedefaultseppunct}\relax
\EndOfBibitem
\bibitem[Chen and Ferguson(2018)Chen, and Ferguson]{chen2018molecular}
Chen,~W.; Ferguson,~A.~L. Molecular enhanced sampling with autoencoders:
  On-the-fly collective variable discovery and accelerated free energy
  landscape exploration. \emph{Journal of computational chemistry}
  \textbf{2018}, \emph{39}, 2079--2102\relax
\mciteBstWouldAddEndPuncttrue
\mciteSetBstMidEndSepPunct{\mcitedefaultmidpunct}
{\mcitedefaultendpunct}{\mcitedefaultseppunct}\relax
\EndOfBibitem
\bibitem[Peters \latin{et~al.}(2007)Peters, Beckham, and
  Trout]{peters2007extensions}
Peters,~B.; Beckham,~G.~T.; Trout,~B.~L. Extensions to the likelihood
  maximization approach for finding reaction coordinates. \emph{The Journal of
  chemical physics} \textbf{2007}, \emph{127}, 034109\relax
\mciteBstWouldAddEndPuncttrue
\mciteSetBstMidEndSepPunct{\mcitedefaultmidpunct}
{\mcitedefaultendpunct}{\mcitedefaultseppunct}\relax
\EndOfBibitem
\bibitem[Sultan and Pande(2018)Sultan, and Pande]{sultan2018automated}
Sultan,~M.~M.; Pande,~V.~S. Automated design of collective variables using
  supervised machine learning. \emph{The Journal of chemical physics}
  \textbf{2018}, \emph{149}, 094106\relax
\mciteBstWouldAddEndPuncttrue
\mciteSetBstMidEndSepPunct{\mcitedefaultmidpunct}
{\mcitedefaultendpunct}{\mcitedefaultseppunct}\relax
\EndOfBibitem
\bibitem[Mendels \latin{et~al.}(2018)Mendels, Piccini, and
  Parrinello]{mendels2018collective}
Mendels,~D.; Piccini,~G.; Parrinello,~M. Collective variables from local
  fluctuations. \emph{The journal of physical chemistry letters} \textbf{2018},
  \emph{9}, 2776--2781\relax
\mciteBstWouldAddEndPuncttrue
\mciteSetBstMidEndSepPunct{\mcitedefaultmidpunct}
{\mcitedefaultendpunct}{\mcitedefaultseppunct}\relax
\EndOfBibitem
\bibitem[Bonati \latin{et~al.}(2020)Bonati, Rizzi, and
  Parrinello]{bonati2020data}
Bonati,~L.; Rizzi,~V.; Parrinello,~M. Data-driven collective variables for
  enhanced sampling. \emph{The journal of physical chemistry letters}
  \textbf{2020}, \emph{11}, 2998--3004\relax
\mciteBstWouldAddEndPuncttrue
\mciteSetBstMidEndSepPunct{\mcitedefaultmidpunct}
{\mcitedefaultendpunct}{\mcitedefaultseppunct}\relax
\EndOfBibitem
\bibitem[Welling(2005)]{welling2005fisher}
Welling,~M. Fisher linear discriminant analysis. \emph{Department of Computer
  Science, University of Toronto} \textbf{2005}, \relax
\mciteBstWouldAddEndPunctfalse
\mciteSetBstMidEndSepPunct{\mcitedefaultmidpunct}
{}{\mcitedefaultseppunct}\relax
\EndOfBibitem
\bibitem[Piccini \latin{et~al.}(2018)Piccini, Mendels, and
  Parrinello]{piccini2018metadynamics}
Piccini,~G.; Mendels,~D.; Parrinello,~M. Metadynamics with discriminants: A
  tool for understanding chemistry. \emph{Journal of chemical theory and
  computation} \textbf{2018}, \emph{14}, 5040--5044\relax
\mciteBstWouldAddEndPuncttrue
\mciteSetBstMidEndSepPunct{\mcitedefaultmidpunct}
{\mcitedefaultendpunct}{\mcitedefaultseppunct}\relax
\EndOfBibitem
\bibitem[Dorfer \latin{et~al.}(2016)Dorfer, Kelz, and Widmer]{dorfer2015deep}
Dorfer,~M.; Kelz,~R.; Widmer,~G. Deep linear discriminant analysis. \emph{4th
  International Conference on Learning Representation ICLR} \textbf{2016},
  \relax
\mciteBstWouldAddEndPunctfalse
\mciteSetBstMidEndSepPunct{\mcitedefaultmidpunct}
{}{\mcitedefaultseppunct}\relax
\EndOfBibitem
\bibitem[Karmakar \latin{et~al.}(2021)Karmakar, Invernizzi, Rizzi, and
  Parrinello]{karmakar2021colvarCryst}
Karmakar,~T.; Invernizzi,~M.; Rizzi,~V.; Parrinello,~M. Collective variables
  for the study of crystallisation. \emph{Molecular Physics} \textbf{2021},
  \emph{0}, e1893848\relax
\mciteBstWouldAddEndPuncttrue
\mciteSetBstMidEndSepPunct{\mcitedefaultmidpunct}
{\mcitedefaultendpunct}{\mcitedefaultseppunct}\relax
\EndOfBibitem
\bibitem[Rizzi \latin{et~al.}(2021)Rizzi, Bonati, Ansari, and
  Parrinello]{rizzi2021role}
Rizzi,~V.; Bonati,~L.; Ansari,~N.; Parrinello,~M. The role of water in
  host-guest interaction. \emph{Nature Communications} \textbf{2021},
  \emph{12}, 1--7\relax
\mciteBstWouldAddEndPuncttrue
\mciteSetBstMidEndSepPunct{\mcitedefaultmidpunct}
{\mcitedefaultendpunct}{\mcitedefaultseppunct}\relax
\EndOfBibitem
\bibitem[Debnath and Parrinello(2020)Debnath, and
  Parrinello]{debnath2020gaussian}
Debnath,~J.; Parrinello,~M. Gaussian mixture-based enhanced sampling for
  statics and dynamics. \emph{The Journal of Physical Chemistry Letters}
  \textbf{2020}, \emph{11}, 5076--5080\relax
\mciteBstWouldAddEndPuncttrue
\mciteSetBstMidEndSepPunct{\mcitedefaultmidpunct}
{\mcitedefaultendpunct}{\mcitedefaultseppunct}\relax
\EndOfBibitem
\bibitem[Rizzi \latin{et~al.}(2019)Rizzi, Mendels, Sicilia, and
  Parrinello]{rizzi2019blind}
Rizzi,~V.; Mendels,~D.; Sicilia,~E.; Parrinello,~M. Blind search for complex
  chemical pathways using harmonic linear discriminant analysis. \emph{Journal
  of chemical theory and computation} \textbf{2019}, \emph{15},
  4507--4515\relax
\mciteBstWouldAddEndPuncttrue
\mciteSetBstMidEndSepPunct{\mcitedefaultmidpunct}
{\mcitedefaultendpunct}{\mcitedefaultseppunct}\relax
\EndOfBibitem
\bibitem[Rumpel and Limbach(1989)Rumpel, and Limbach]{rumpel1989nmr}
Rumpel,~H.; Limbach,~H.~H. NMR study of kinetic HH/HD/DD isotope, solvent and
  solid-state effects on the double proton transfer in azophenine.
  \emph{Journal of the American Chemical Society} \textbf{1989}, \emph{111},
  5429--5441\relax
\mciteBstWouldAddEndPuncttrue
\mciteSetBstMidEndSepPunct{\mcitedefaultmidpunct}
{\mcitedefaultendpunct}{\mcitedefaultseppunct}\relax
\EndOfBibitem
\bibitem[Fatollahpour and Tahermansouri(2017)Fatollahpour, and
  Tahermansouri]{fatollahpour2017dft}
Fatollahpour,~M.; Tahermansouri,~H. DFT study of the intramolecular double
  proton transfer of 2, 5-diamino-1, 4-benzoquinone and its derivatives, and
  investigations about their aromaticity. \emph{Comptes Rendus Chimie}
  \textbf{2017}, \emph{20}, 942--951\relax
\mciteBstWouldAddEndPuncttrue
\mciteSetBstMidEndSepPunct{\mcitedefaultmidpunct}
{\mcitedefaultendpunct}{\mcitedefaultseppunct}\relax
\EndOfBibitem
\bibitem[Bonomi \latin{et~al.}(2009)Bonomi, Branduardi, Bussi, Camilloni,
  Provasi, Raiteri, Donadio, Marinelli, Pietrucci, Broglia, \latin{et~al.}
  others]{bonomi2009plumed}
Bonomi,~M.; Branduardi,~D.; Bussi,~G.; Camilloni,~C.; Provasi,~D.; Raiteri,~P.;
  Donadio,~D.; Marinelli,~F.; Pietrucci,~F.; Broglia,~R.~A. \latin{et~al.}
  PLUMED: A portable plugin for free-energy calculations with molecular
  dynamics. \emph{Computer Physics Communications} \textbf{2009}, \emph{180},
  1961--1972\relax
\mciteBstWouldAddEndPuncttrue
\mciteSetBstMidEndSepPunct{\mcitedefaultmidpunct}
{\mcitedefaultendpunct}{\mcitedefaultseppunct}\relax
\EndOfBibitem
\bibitem[Tribello \latin{et~al.}(2014)Tribello, Bonomi, Branduardi, Camilloni,
  and Bussi]{tribello2014plumed}
Tribello,~G.~A.; Bonomi,~M.; Branduardi,~D.; Camilloni,~C.; Bussi,~G. PLUMED 2:
  New feathers for an old bird. \emph{Computer Physics Communications}
  \textbf{2014}, \emph{185}, 604--613\relax
\mciteBstWouldAddEndPuncttrue
\mciteSetBstMidEndSepPunct{\mcitedefaultmidpunct}
{\mcitedefaultendpunct}{\mcitedefaultseppunct}\relax
\EndOfBibitem
\bibitem[Bonomi \latin{et~al.}(2019)Bonomi, Bussi, Camilloni, Tribello,
  Ban{\'a}{\v{s}}, Barducci, Bernetti, Bolhuis, Bottaro, Branduardi,
  \latin{et~al.} others]{bonomi2019promoting}
Bonomi,~M.; Bussi,~G.; Camilloni,~C.; Tribello,~G.~A.; Ban{\'a}{\v{s}},~P.;
  Barducci,~A.; Bernetti,~M.; Bolhuis,~P.~G.; Bottaro,~S.; Branduardi,~D.
  \latin{et~al.}  Promoting transparency and reproducibility in enhanced
  molecular simulations. \emph{Nature methods} \textbf{2019}, \emph{16},
  670--673\relax
\mciteBstWouldAddEndPuncttrue
\mciteSetBstMidEndSepPunct{\mcitedefaultmidpunct}
{\mcitedefaultendpunct}{\mcitedefaultseppunct}\relax
\EndOfBibitem
\bibitem[Abraham \latin{et~al.}(2015)Abraham, Murtola, Schulz, P{\'a}ll, Smith,
  Hess, and Lindahl]{abraham2015gromacs}
Abraham,~M.~J.; Murtola,~T.; Schulz,~R.; P{\'a}ll,~S.; Smith,~J.~C.; Hess,~B.;
  Lindahl,~E. GROMACS: High performance molecular simulations through
  multi-level parallelism from laptops to supercomputers. \emph{SoftwareX}
  \textbf{2015}, \emph{1}, 19--25\relax
\mciteBstWouldAddEndPuncttrue
\mciteSetBstMidEndSepPunct{\mcitedefaultmidpunct}
{\mcitedefaultendpunct}{\mcitedefaultseppunct}\relax
\EndOfBibitem
\bibitem[K{\"u}hne \latin{et~al.}(2020)K{\"u}hne, Iannuzzi, Del~Ben, Rybkin,
  Seewald, Stein, Laino, Khaliullin, Sch{\"u}tt, Schiffmann, \latin{et~al.}
  others]{kuhne2020cp2k}
K{\"u}hne,~T.~D.; Iannuzzi,~M.; Del~Ben,~M.; Rybkin,~V.~V.; Seewald,~P.;
  Stein,~F.; Laino,~T.; Khaliullin,~R.~Z.; Sch{\"u}tt,~O.; Schiffmann,~F.
  \latin{et~al.}  CP2K: An electronic structure and molecular dynamics software
  package-Quickstep: Efficient and accurate electronic structure calculations.
  \emph{The Journal of Chemical Physics} \textbf{2020}, \emph{152},
  194103\relax
\mciteBstWouldAddEndPuncttrue
\mciteSetBstMidEndSepPunct{\mcitedefaultmidpunct}
{\mcitedefaultendpunct}{\mcitedefaultseppunct}\relax
\EndOfBibitem
\bibitem[Paszke \latin{et~al.}(2017)Paszke, Gross, Chintala, Chanan, Yang,
  DeVito, Lin, Desmaison, Antiga, and Lerer]{paszke2017automatic}
Paszke,~A.; Gross,~S.; Chintala,~S.; Chanan,~G.; Yang,~E.; DeVito,~Z.; Lin,~Z.;
  Desmaison,~A.; Antiga,~L.; Lerer,~A. Automatic differentiation in pytorch.
  \emph{Advances in Neural Information Processing Systems} \textbf{2017},
  \relax
\mciteBstWouldAddEndPunctfalse
\mciteSetBstMidEndSepPunct{\mcitedefaultmidpunct}
{}{\mcitedefaultseppunct}\relax
\EndOfBibitem
\end{mcitethebibliography}


\providecommand{\latin}[1]{#1}
\makeatletter
\providecommand{\doi}
  {\begingroup\let\do\@makeother\dospecials
  \catcode`\{=1 \catcode`\}=2 \doi@aux}
\providecommand{\doi@aux}[1]{\endgroup\texttt{#1}}
\makeatother
\providecommand*\mcitethebibliography{\thebibliography}
\csname @ifundefined\endcsname{endmcitethebibliography}
  {\let\endmcitethebibliography\endthebibliography}{}
\begin{mcitethebibliography}{17}
\providecommand*\natexlab[1]{#1}
\providecommand*\mciteSetBstSublistMode[1]{}
\providecommand*\mciteSetBstMaxWidthForm[2]{}
\providecommand*\mciteBstWouldAddEndPuncttrue
  {\def\EndOfBibitem{\unskip.}}
\providecommand*\mciteBstWouldAddEndPunctfalse
  {\let\EndOfBibitem\relax}
\providecommand*\mciteSetBstMidEndSepPunct[3]{}
\providecommand*\mciteSetBstSublistLabelBeginEnd[3]{}
\providecommand*\EndOfBibitem{}
\mciteSetBstSublistMode{f}
\mciteSetBstMaxWidthForm{subitem}{(\alph{mcitesubitemcount})}
\mciteSetBstSublistLabelBeginEnd
  {\mcitemaxwidthsubitemform\space}
  {\relax}
  {\relax}

\bibitem[Bonati \latin{et~al.}(2020)Bonati, Rizzi, and
  Parrinello]{bonati2020data}
Bonati,~L.; Rizzi,~V.; Parrinello,~M. Data-driven collective variables for
  enhanced sampling. \emph{The journal of physical chemistry letters}
  \textbf{2020}, \emph{11}, 2998--3004\relax
\mciteBstWouldAddEndPuncttrue
\mciteSetBstMidEndSepPunct{\mcitedefaultmidpunct}
{\mcitedefaultendpunct}{\mcitedefaultseppunct}\relax
\EndOfBibitem
\bibitem[Welling(2005)]{welling2005fisher}
Welling,~M. Fisher linear discriminant analysis. \emph{Department of Computer
  Science, University of Toronto} \textbf{2005}, \relax
\mciteBstWouldAddEndPunctfalse
\mciteSetBstMidEndSepPunct{\mcitedefaultmidpunct}
{}{\mcitedefaultseppunct}\relax
\EndOfBibitem
\bibitem[Paszke \latin{et~al.}(2017)Paszke, Gross, Chintala, Chanan, Yang,
  DeVito, Lin, Desmaison, Antiga, and Lerer]{paszke2017automatic}
Paszke,~A.; Gross,~S.; Chintala,~S.; Chanan,~G.; Yang,~E.; DeVito,~Z.; Lin,~Z.;
  Desmaison,~A.; Antiga,~L.; Lerer,~A. Automatic differentiation in pytorch.
  \emph{Advances in Neural Information Processing Systems} \textbf{2017},
  \relax
\mciteBstWouldAddEndPunctfalse
\mciteSetBstMidEndSepPunct{\mcitedefaultmidpunct}
{}{\mcitedefaultseppunct}\relax
\EndOfBibitem
\bibitem[Kingma and Ba(2014)Kingma, and Ba]{kingma2014adam}
Kingma,~D.~P.; Ba,~J. Adam: A method for stochastic optimization. \emph{3rd
  ICLR} \textbf{2014}, \relax
\mciteBstWouldAddEndPunctfalse
\mciteSetBstMidEndSepPunct{\mcitedefaultmidpunct}
{}{\mcitedefaultseppunct}\relax
\EndOfBibitem
\bibitem[Bonomi \latin{et~al.}(2019)Bonomi, Bussi, Camilloni, Tribello,
  Ban{\'a}{\v{s}}, Barducci, Bernetti, Bolhuis, Bottaro, Branduardi,
  \latin{et~al.} others]{bonomi2019promoting}
Bonomi,~M.; Bussi,~G.; Camilloni,~C.; Tribello,~G.~A.; Ban{\'a}{\v{s}},~P.;
  Barducci,~A.; Bernetti,~M.; Bolhuis,~P.~G.; Bottaro,~S.; Branduardi,~D.,
  \latin{et~al.}  Promoting transparency and reproducibility in enhanced
  molecular simulations. \emph{Nature methods} \textbf{2019}, \emph{16},
  670--673\relax
\mciteBstWouldAddEndPuncttrue
\mciteSetBstMidEndSepPunct{\mcitedefaultmidpunct}
{\mcitedefaultendpunct}{\mcitedefaultseppunct}\relax
\EndOfBibitem
\bibitem[Abraham \latin{et~al.}(2015)Abraham, Murtola, Schulz, P{\'a}ll, Smith,
  Hess, and Lindahl]{abraham2015gromacs}
Abraham,~M.~J.; Murtola,~T.; Schulz,~R.; P{\'a}ll,~S.; Smith,~J.~C.; Hess,~B.;
  Lindahl,~E. GROMACS: High performance molecular simulations through
  multi-level parallelism from laptops to supercomputers. \emph{SoftwareX}
  \textbf{2015}, \emph{1}, 19--25\relax
\mciteBstWouldAddEndPuncttrue
\mciteSetBstMidEndSepPunct{\mcitedefaultmidpunct}
{\mcitedefaultendpunct}{\mcitedefaultseppunct}\relax
\EndOfBibitem
\bibitem[Hornak \latin{et~al.}(2006)Hornak, Abel, Okur, Strockbine, Roitberg,
  and Simmerling]{hornak2006comparison}
Hornak,~V.; Abel,~R.; Okur,~A.; Strockbine,~B.; Roitberg,~A.; Simmerling,~C.
  Comparison of multiple Amber force fields and development of improved protein
  backbone parameters. \emph{Proteins: Structure, Function, and Bioinformatics}
  \textbf{2006}, \emph{65}, 712--725\relax
\mciteBstWouldAddEndPuncttrue
\mciteSetBstMidEndSepPunct{\mcitedefaultmidpunct}
{\mcitedefaultendpunct}{\mcitedefaultseppunct}\relax
\EndOfBibitem
\bibitem[Bussi \latin{et~al.}(2007)Bussi, Donadio, and
  Parrinello]{bussi2007canonical}
Bussi,~G.; Donadio,~D.; Parrinello,~M. Canonical sampling through velocity
  rescaling. \emph{The Journal of chemical physics} \textbf{2007}, \emph{126},
  014101\relax
\mciteBstWouldAddEndPuncttrue
\mciteSetBstMidEndSepPunct{\mcitedefaultmidpunct}
{\mcitedefaultendpunct}{\mcitedefaultseppunct}\relax
\EndOfBibitem
\bibitem[Yin \latin{et~al.}(2017)Yin, Henriksen, Slochower, Shirts, Chiu,
  Mobley, and Gilson]{yin2017overview}
Yin,~J.; Henriksen,~N.~M.; Slochower,~D.~R.; Shirts,~M.~R.; Chiu,~M.~W.;
  Mobley,~D.~L.; Gilson,~M.~K. Overview of the SAMPL5 host--guest challenge:
  Are we doing better? \emph{Journal of computer-aided molecular design}
  \textbf{2017}, \emph{31}, 1--19\relax
\mciteBstWouldAddEndPuncttrue
\mciteSetBstMidEndSepPunct{\mcitedefaultmidpunct}
{\mcitedefaultendpunct}{\mcitedefaultseppunct}\relax
\EndOfBibitem
\bibitem[Rizzi \latin{et~al.}(2021)Rizzi, Bonati, Ansari, and
  Parrinello]{rizzi2021role}
Rizzi,~V.; Bonati,~L.; Ansari,~N.; Parrinello,~M. The role of water in
  host-guest interaction. \emph{Nature Communications} \textbf{2021},
  \emph{12}, 1--7\relax
\mciteBstWouldAddEndPuncttrue
\mciteSetBstMidEndSepPunct{\mcitedefaultmidpunct}
{\mcitedefaultendpunct}{\mcitedefaultseppunct}\relax
\EndOfBibitem
\bibitem[Limongelli \latin{et~al.}(2013)Limongelli, Bonomi, and
  Parrinello]{limongelli2013funnel}
Limongelli,~V.; Bonomi,~M.; Parrinello,~M. Funnel metadynamics as accurate
  binding free-energy method. \emph{Proceedings of the National Academy of
  Sciences} \textbf{2013}, \emph{110}, 6358--6363\relax
\mciteBstWouldAddEndPuncttrue
\mciteSetBstMidEndSepPunct{\mcitedefaultmidpunct}
{\mcitedefaultendpunct}{\mcitedefaultseppunct}\relax
\EndOfBibitem
\bibitem[Wang \latin{et~al.}(2004)Wang, Wolf, Caldwell, Kollman, and
  Case]{wang2004development}
Wang,~J.; Wolf,~R.~M.; Caldwell,~J.~W.; Kollman,~P.~A.; Case,~D.~A. Development
  and testing of a general amber force field. \emph{Journal of computational
  chemistry} \textbf{2004}, \emph{25}, 1157--1174\relax
\mciteBstWouldAddEndPuncttrue
\mciteSetBstMidEndSepPunct{\mcitedefaultmidpunct}
{\mcitedefaultendpunct}{\mcitedefaultseppunct}\relax
\EndOfBibitem
\bibitem[Bayly \latin{et~al.}(1993)Bayly, Cieplak, Cornell, and
  Kollman]{bayly1993well}
Bayly,~C.~I.; Cieplak,~P.; Cornell,~W.; Kollman,~P.~A. A well-behaved
  electrostatic potential based method using charge restraints for deriving
  atomic charges: the RESP model. \emph{The Journal of Physical Chemistry}
  \textbf{1993}, \emph{97}, 10269--10280\relax
\mciteBstWouldAddEndPuncttrue
\mciteSetBstMidEndSepPunct{\mcitedefaultmidpunct}
{\mcitedefaultendpunct}{\mcitedefaultseppunct}\relax
\EndOfBibitem
\bibitem[Jorgensen \latin{et~al.}(1983)Jorgensen, Chandrasekhar, Madura, Impey,
  and Klein]{jorgensen1983comparison}
Jorgensen,~W.~L.; Chandrasekhar,~J.; Madura,~J.~D.; Impey,~R.~W.; Klein,~M.~L.
  Comparison of simple potential functions for simulating liquid water.
  \emph{The Journal of chemical physics} \textbf{1983}, \emph{79},
  926--935\relax
\mciteBstWouldAddEndPuncttrue
\mciteSetBstMidEndSepPunct{\mcitedefaultmidpunct}
{\mcitedefaultendpunct}{\mcitedefaultseppunct}\relax
\EndOfBibitem
\bibitem[K{\"u}hne \latin{et~al.}(2020)K{\"u}hne, Iannuzzi, Del~Ben, Rybkin,
  Seewald, Stein, Laino, Khaliullin, Sch{\"u}tt, Schiffmann, \latin{et~al.}
  others]{kuhne2020cp2k}
K{\"u}hne,~T.~D.; Iannuzzi,~M.; Del~Ben,~M.; Rybkin,~V.~V.; Seewald,~P.;
  Stein,~F.; Laino,~T.; Khaliullin,~R.~Z.; Sch{\"u}tt,~O.; Schiffmann,~F.,
  \latin{et~al.}  CP2K: An electronic structure and molecular dynamics software
  package-Quickstep: Efficient and accurate electronic structure calculations.
  \emph{The Journal of Chemical Physics} \textbf{2020}, \emph{152},
  194103\relax
\mciteBstWouldAddEndPuncttrue
\mciteSetBstMidEndSepPunct{\mcitedefaultmidpunct}
{\mcitedefaultendpunct}{\mcitedefaultseppunct}\relax
\EndOfBibitem
\bibitem[Debnath and Parrinello(2020)Debnath, and
  Parrinello]{debnath2020gaussian}
Debnath,~J.; Parrinello,~M. Gaussian mixture-based enhanced sampling for
  statics and dynamics. \emph{The Journal of Physical Chemistry Letters}
  \textbf{2020}, \emph{11}, 5076--5080\relax
\mciteBstWouldAddEndPuncttrue
\mciteSetBstMidEndSepPunct{\mcitedefaultmidpunct}
{\mcitedefaultendpunct}{\mcitedefaultseppunct}\relax
\EndOfBibitem
\end{mcitethebibliography}

\end{document}


\section{Deep-DA Model - Methods}
    
    Deep-LDA\cite{bonati2020data}, like ordinary LDA\cite{welling2005fisher}, turns the discrimination analysis into a variational problem whose solution requires a linear step. In Deep-DA approach, we skip this linear step and express $s$ directly as the unique output of a NN.
    For the sake of simplicity, we consider here only the two-state scenario, which is a case commonly encountered in the practice.
    Given two states A and B, characterized by a set of descriptors \textbf{d}, we want to find a projection $s$ along which the two states are well-discriminated.
    As in Fisher's approach, we start from an initial guess for s and we calculate the average values $\mu_A$ and $\mu_B$ that the data in the two states have in the projected space. Similarly we evaluate the variances $\sigma_A^2$ and $\sigma_B^2$.
    Then we measure the ability of s to discriminate states A and B with a Fisher's ratio
            \begin{equation}
                F = \frac{(\mu_A - \mu_B)^2}{\epsilon + \sigma_A^2 + \sigma_B^2}
                \label{eq:fisher_eps}
            \end{equation}
    in which a regularizing positive $\epsilon$  parameter has been added to the denominator. 
    The optimal projection $s$ is found by maximizing $F$. 

    However, if we leave complete variational flexibility to $s$, $F$ is not bound and eventually diverges  as  $(\mu_A - \mu_B)^2 \to \infty$. Thus we stop the NN training when a preassigned  $\Delta = \sqrt{F}$ value is reached.  Note that the $\epsilon$ in the denominator prevents the denominator from going to zero  and  controls  the final width of the distributions. For the choice of $\Delta$ the same prescription given in the main text can be followed.

\section{Neural Network training and architecture}

    All the models have been trained using the Pytorch\cite{paszke2017automatic} library. A Google Colab notebook tutorial on how to train a Deep TDA CV can be found in \href{https://colab.research.google.com/drive/1TO7bAkmIznsdfea2i5NXfNtytJrnkkIt?usp=sharing}{Deep-TDA tutorial}\footnote{https://colab.research.google.com/drive/1TO7bAkmIznsdfea2i5NXfNtytJrnkkIt?usp=sharing}.

    The Neural Networks (NNs) have been trained on $10^4$ configurations taken from each metastable basin. Two thirds of the shuffled dataset were used for training, typically divided in two batches. The NN was then validated on the remaining one third.
    The training data have been standardized to have values within -1 and 1.
    
    Following Ref.\cite{bonati2020data}, the number of layers in the NN has been kept the same for every system, consisting in the input layer and three hidden layers. The number of nodes per layer also was similar to what done in Ref.\cite{bonati2020data}. However, some adjustments have been made to optimize the performance for the various systems and are specified below.
    
    We have used the Rectified Linear Unit (ReLU) activation function and the ADAM\cite{kingma2014adam} optimizer with a learning rate of $10^{-3}$.
    The loss functions for the two models are given in Eqs.\ref{eq:fisher},\ref{eq:loss},\ref{eq:loss_multi} and we have applied a L2 regularization to the weights $\Theta_i$ of $10^{-5}$.
    
    \paragraph{PLUMED interface with Pytorch}
        To collect the input data for the NN and to perform enhanced sampling with the trained CVs we have used the open-source plugin PLUMED2\cite{bonomi2019promoting}.
        To import the trained model from Pytorch to PLUMED we have used the interface developed by Luigi Bonati\cite{bonati2020data}. This is developed on the Python C++ API Libtorch and is available on \href{https://github.com/luigibonati/data-driven-CVs/}{Github}\footnote{https://github.com/luigibonati/data-driven-CVs/}.

\section{Alanine Dipeptide}\label{sec:S_alanine}

    \paragraph{General computational set up}
        The alanine dipeptide simulations have been carried out using the GROMACS-2019.6\cite{abraham2015gromacs}  MD engine patched with PLUMED2.7. We have used the Amber99-SB\cite{hornak2006comparison} force field with a 2fs timestep.
        To sample the NVT ensemble we have used the velocity rescaling thermostat \cite{bussi2007canonical} set at 300K.
    
    \subsection{Alanine Dipeptide - Deep-TDA model}
        In the OPES simulations driven by the Deep-TDA CV, we have used \texttt{SIGMA=0.2} to match the standard deviation of the basins we had set in the target distribution, \texttt{BARRIER=30} kJ/mol and \texttt{PACE=500}. 
        We used the 45 distances between heavy atoms as input of a NN with three hidden layers \{24, 12, 1\}. The target function parameters were $\mu^{tg}_A = -7$, $\mu^{tg}_B = 7$ and $\sigma^{tg}=0.2$. The resulting $\Delta$ separation was thus $\Delta=50$.
        The hyperparameters for the loss function (Eq.\ref{eq:loss}) were $\alpha=1$ and $\beta=250$.
        We performed 5 statistically independent 10ns simulations to compute the average values reported in the main text.
        
    \subsection{Alanine Dipeptide - Deep-DA model}
        In the OPES simulations with the Deep-DA CV we have used \texttt{SIGMA=0.2} to match the smallest standard deviation of the two training basins, \texttt{BARRIER=30} kJ/mol and \texttt{PACE=500}. We used the 45 distances between heavy atoms as input of a NN with three hidden layers \{25, 10, 1\} and the training was stopped at the threshold $\Delta=42$.
        We averaged over 4 statistically independent 10ns simulations.
        
        \begin{figure}[h!]
            \begin{minipage}{0.49\linewidth}
            \includegraphics[width=\linewidth]{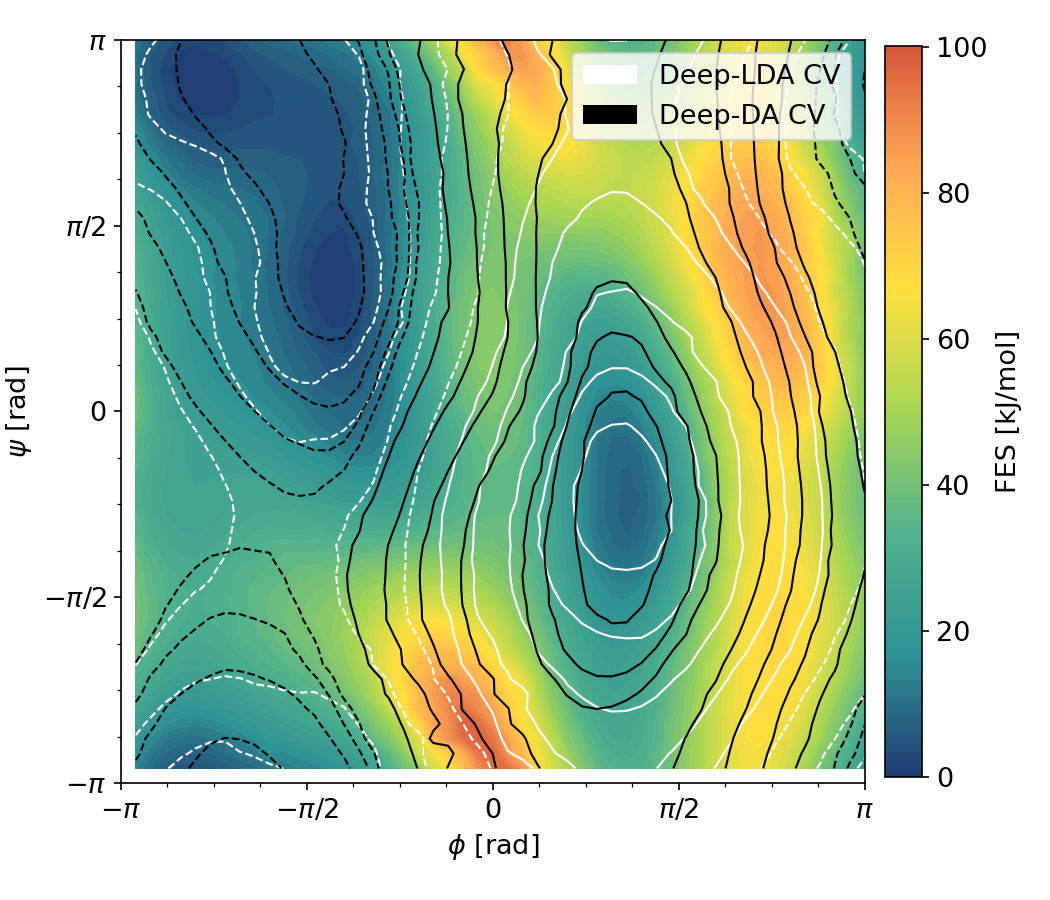}
            \end{minipage}
            \hspace*{\fill}
            \begin{minipage}{0.49\linewidth}
            \includegraphics[width=\linewidth]{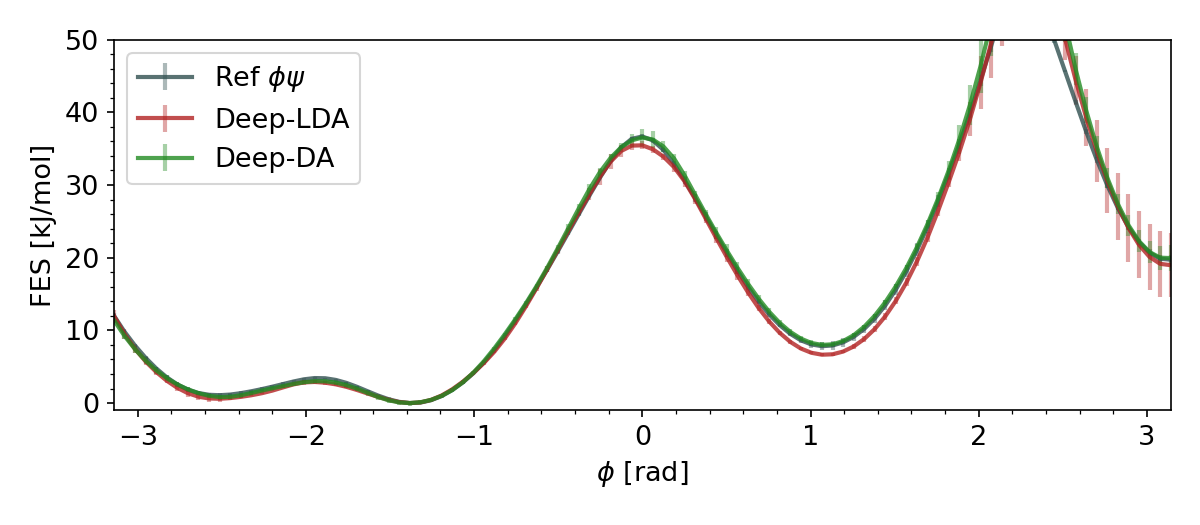}
            \includegraphics[width=\linewidth]{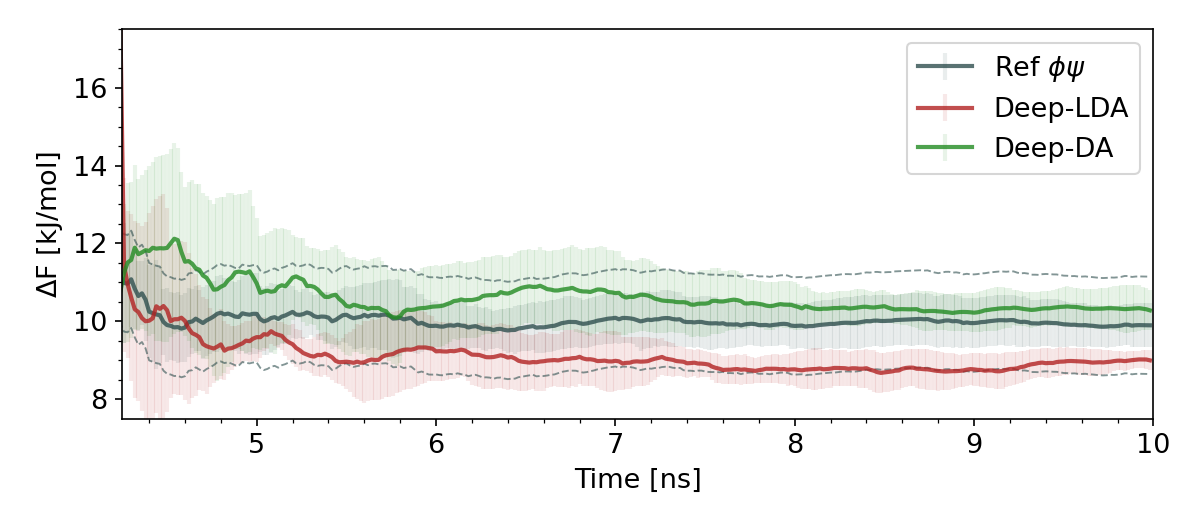}
            \end{minipage}
            
            \caption{Results of OPES enhanced simulations of the folding of alanine dipeptide. Left: Comparison of the isolines of Deep-LDA (white) and Deep-DA (black) CVs on top of the energetic landscape in Ramachandran angles $\phi\psi$ plane. Solid lines are for positive CV values, dashed lines for negative ones.
            Right: Comparison of energy estimates from OPES simulations using Deep-LDA, Deep-DA or a reference $\phi\psi$ as CVs. Top-Right: FES profile estimates along the $\phi$ Ramachandran angle. Bottom-Right: $\Delta$F between the metastable basins estimates obtained as functions of the simulation time. The dashed lines give the $\pm0.5K_bT$ range on the reference curve.}
            \label{fig:S_alanine}
        \end{figure}
        In Fig.\ref{fig:S_alanine} are reported for Deep-DA model the same plots given for Deep-TDA model in Fig.\ref{fig:alanine} of the main text. These compare the isolines of Deep-LDA and Deep-DA on the energy landscape in the Ramachandran plot (left panel), the FES estimates with the reference of the Ramachandran angles used as CVs (top-right panel) and the estimate on the $\Delta F$ between the basins (bottom-right panel). Also with this model the results are almost equivalent on all these fronts.

\section{Calixarene host - G2 guest system}

        We also tested the method on a much-studied model host-guest system that is part of the SAMPL5\cite{yin2017overview} challenge. This consists of a calyxarene host (OAMe) and a small planar ligand (G2) solvated in water (see Fig.\ref{fig:calixarene}). This system has also been studied in our group and it was found out that good binding free energy values for the ligand could be obtained if one used as CVs the vertical distance from the binding site z and a CV that describes water solvation in the cavity and around the ligand. The latter is the most innovative aspect of Ref.\cite{rizzi2021role} and was obtained by using Deep-LDA. The descriptors used as input of the Deep-LDA NN report on water solvation around carefully selected points, either fixed in space or attached to the ligand. These are water coordination numbers (CNs) with respect to 12 points: 4 L points on heavy atoms of the ligand and 8 V points along the vertical axis $z$ of the calixarene. The V points start at the center of the lower phenyl rings of the host and go upwards with a spacing of 2.5{\AA}.
        A detailed description of the descriptors and of the funnel restraint correction can be found in Ref.\cite{rizzi2021role}.

        Here, Deep-TDA is used to redesign a CV that accounts for the role of solvent molecules in the same fashion as done with Deep-LDA in Ref.\cite{rizzi2021role}.
        In order to prevent the ligand from wandering aimlessly in the solution once out of the binding site, we have also applied the same funnel restraint potential used in Ref.\cite{limongelli2013funnel} and at the end corrected the binding energy for the entropy reduction that the funnel constraint entails.

        The training data were collected in unbiased MD runs in the bound and unbound states. 
        To enhance the binding process, we used OPES with our solvation describing Deep-TDA CV and the position of the ligand z.

    \paragraph{General computational set up}
        For the calixarene host-guest system in explicit solvent we have adopted the same parameters and set of descriptors used in Ref.\cite{rizzi2021role}, here we recall just the main points thus we refer the interested reader to the original paper for further details. The simulations have been carried out using GROMACS-2019.6\cite{abraham2015gromacs} patched with PLUMED2.7. We used the GAFF\cite{wang2004development} force field with RESP\cite{bayly1993well} charges and the TIP3P\cite{jorgensen1983comparison} water model.
        
        The timestep was $2$fs and we used the velocity rescaling thermostat set as 300K with a time constant of 100fs.
        
        The sides of the simulation box are 40{\AA} long and it contains 2100 explicit water molecules with the ligand G2 and the host molecule OAMe. Sodium ions are included to impose charge neutrality.

    \subsection{Calixarene host-guest system - Deep-TDA model}
        In the OPES simulations with the Deep-TDA and the $z$ distance CVs we have used the OPES adaptive sigma feature, \texttt{BARRIER=50} kJ/mol and \texttt{PACE=500}.
        
        We used the 12 CN described above as input of a NN with three hidden layers \{24, 12, 1\}. The target function parameters were $\mu^{tg}_A = -4$, $\mu^{tg}_B = 4$ and $\sigma^{tg}=0.2$, giving $\Delta=28$.
        The hyperparameters in the loss function (Eq.\ref{eq:loss}) were $\alpha=1$ and $\beta=250$.
        We averaged over 3 statistically independent 140ns runs.
        
\begin{figure}[h!]
                \centering                  \includegraphics[width=0.6\linewidth]{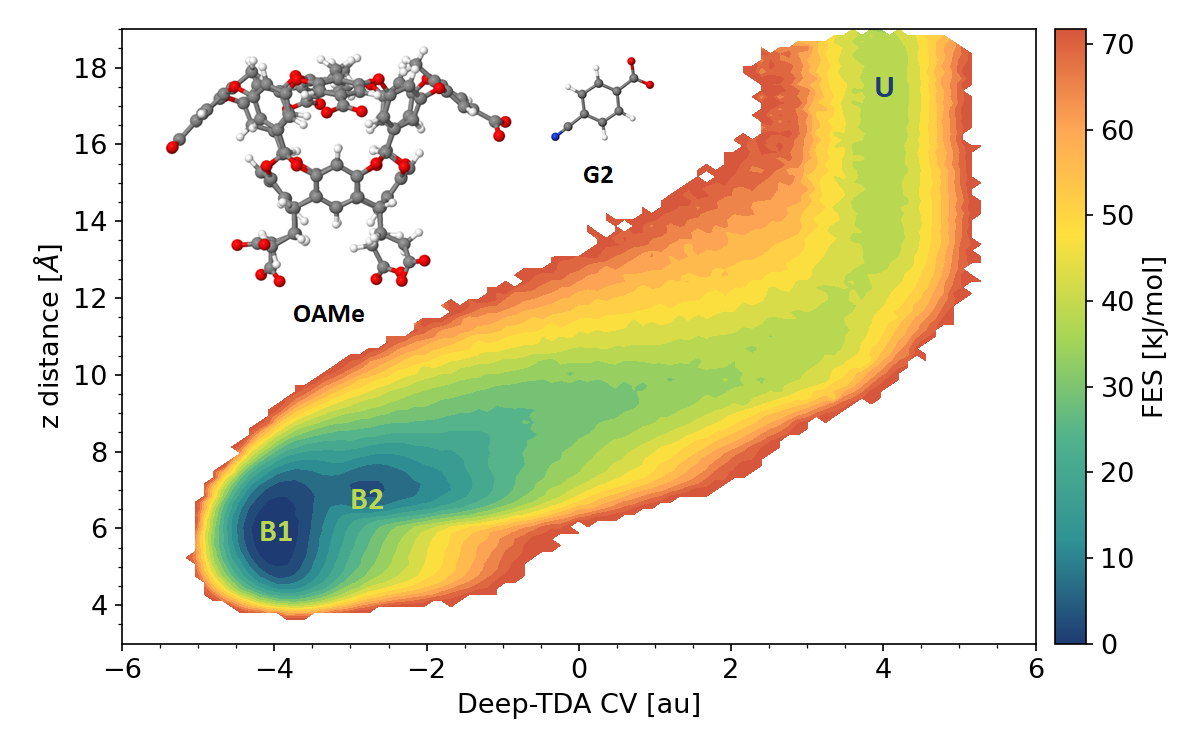}
                \caption{FES estimate in the Deep-TDA CV and z distance plane. The labels indicate the bound states without water in the pocket (B1), with a single water molecule inside (B2) and the unbound state (U).}
                \label{fig:calixarene}
            \end{figure}

        The FES as a function of z distance and Deep-TDA CV is reported in Fig.\ref{fig:calixarene} showing two bound states.
        The B1 is a dry bound state, with only the G2 molecule inside the pocket. On the other hand, the B2 is a wet bound state with a water molecule at the bottom of the pocket just below the ligand.
        
        The results are, within the statistical uncertainty, indistinguishable from those of Ref.\cite{rizzi2021role}. Under the same conditions, the estimates for the binding energy were similar for Deep-LDA, $-22.3\pm1.3$kJ/mol, and Deep-TDA, $-22.3\pm0.7$kJ/mol, and in both cases consistent with the experimental value $-21.1$kJ/mol\cite{yin2017overview}.

    \subsection{Calixarene host-guest system - Deep-DA model}
        In the OPES simulations with the Deep-DA and the $z$ distance CVs we have used the OPES adaptive sigma feature, \texttt{BARRIER=50} kJ/mol and \texttt{PACE=500}.
        We used the 12 CN described above as input of a NN with three hidden layers \{24, 12, 1\} and the training was stopped at the threshold $\Delta=28$. 
        We averaged over 3 statistically independent 140ns simulations.
        
\begin{figure} [h!]
            \centering
            \includegraphics[width=0.6\linewidth]{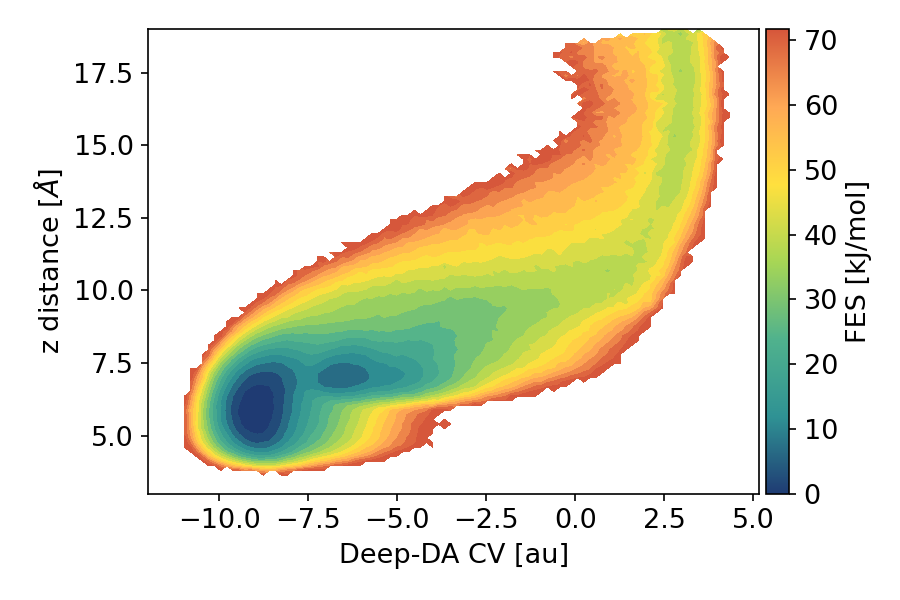}
            \caption{FES estimate for the calixarene host-guest system in the Deep-DA CV and $z$ distance plane. }
            \label{fig:S_calixDA_energy}
        \end{figure}

        The estimates of the FES in the Deep-DA CV and $z$ distance plane are reported in Fig.\ref{fig:S_calixDA_energy}.
        
        The binding energy estimate obtained with Deep-DA is $-22.40\pm0.8$kJ/mol. This value is consistent with the Deep-TDA estimate $-22.3\pm0.7$kJ/mol, the Deep-LDA estimate of $-22.3\pm1.3$kJ/mol and the experimental reference $-22.1$kJ/mol.

\section{Hydrobromination of propene}

    \paragraph{General computational set up}
        The hydrobromination of propene simulations have been carried out using the CP2K-7.1\cite{kuhne2020cp2k} software package patched with PLUMED2.7 at PM6 semi-empirical level. 
        The integration step was $0.5$fs and we used the velocity rescaling thermostat set at 300K with a time constant of 100fs.

\begin{figure} [h!]
            \centering
            \includegraphics[width=0.3\linewidth]{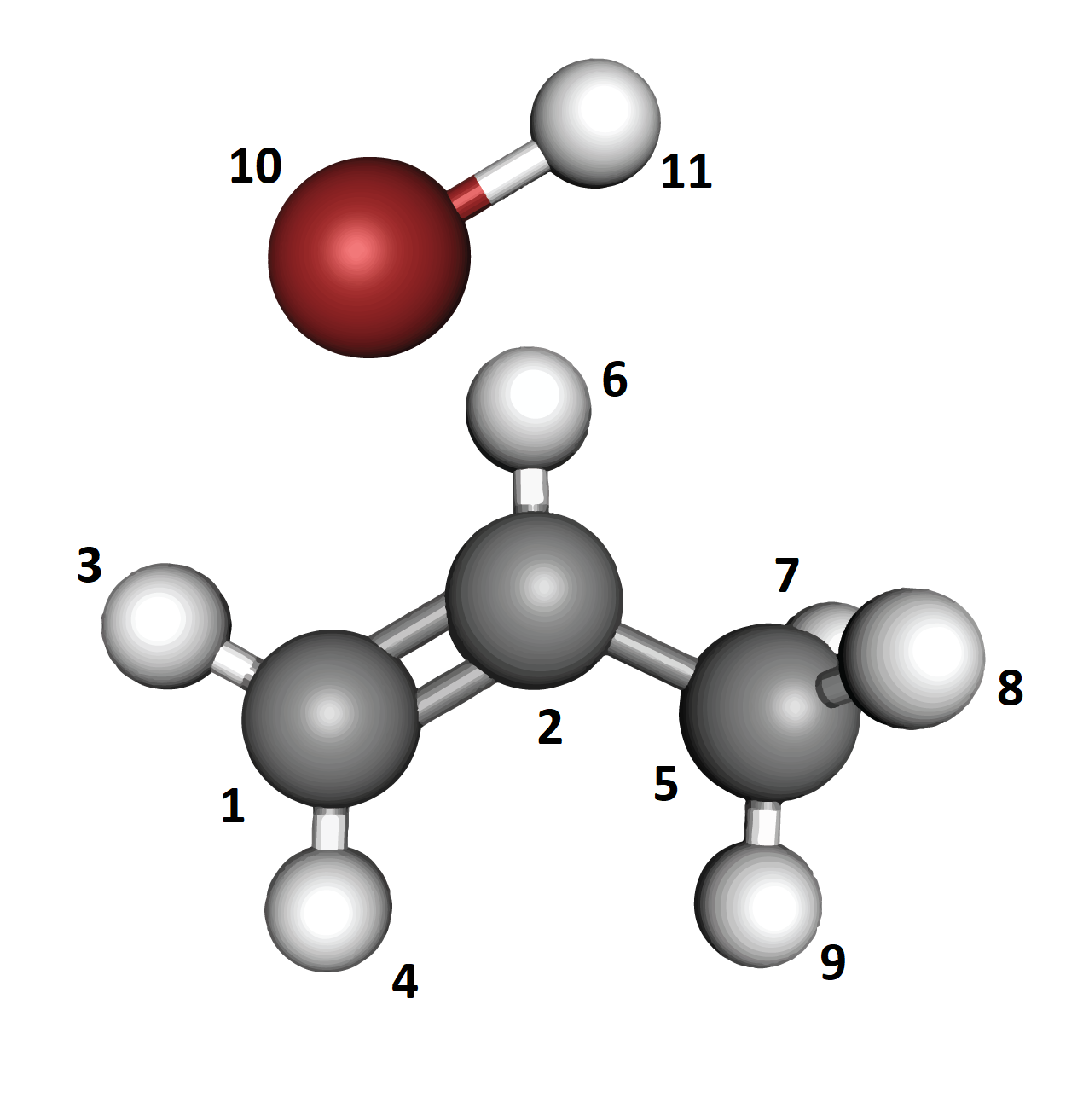}
            \caption{Sketch of the molecules involved in the hydrobromination of propene with the atomic labels.}
            \label{fig:S_hbro}
        \end{figure}

        The contacts used for the training have been computed with PLUMED. The $\sigma_{ij}$ parameters of Eq.\ref{eq:contact} have been:   $\sigma_{BrC}$=1.9{\AA}, $\sigma_{CC}$=1.7{\AA}, $\sigma_{BrH}$=1.4{\AA}, $\sigma_{CH}$=1.2{\AA}. The exponent in the same function have been n=6 and m=8 to have a smooth behavior over a wide range of distances. Contacts between hydrogen atoms have not been used in the training.
        In Fig.\ref{fig:S_hbro} are reported the reagents molecules with the atomic labels used to identify the atoms in PLUMED.
        Following Ref.\cite{debnath2020gaussian}, some restraints have been applied on some of the distances to enforce only $H^{11}$, $B$r, $C^1$ and $C^2$ atoms to take part to the reaction and to prevent undesired reactions to occur:
        \begin{itemize}
            \item To keep the molecules close to each other and in suitable conditions to react, some restraints have been applied on the distances most affected by the reaction ($d_{1-10}$, $d_{1-11}$, $d_{2-10}$, $d_{2-11}$, $d_{10-11}$) in the form $k(d_{ij}-d_0)^2$, with $k=250$KJ/mol and $d_0$=3{\AA}.
        
            \item All the hydrogen-hydrogen distances $d_{ij}$ ($i,j$ in \texttt{ATOMS=3,4,6,7,8,9,11}) have been restrained to be greater than 1.4{\AA} with a restraint in the form $k(d_{ij}-d_0)^2$, with $k=250$kJ/mol and $d_0$=1.4{\AA}.
            
            \item \textit{Non-reactive} hydrogens (\texttt{ATOMS=3,4,6,7,8,9}) have been locked to the respective carbon atoms (\texttt{ATOMS=1,2,5}) with restraints on all the distances $d_{HC}$ in the form $k(d_{HC}-d_0+o_i)^2$, with $k=450$kJ/mol, $d_0$=1.4{\AA} and $o_i$=0.2.
        \end{itemize}
        
    \paragraph{Two-dimensional CV}
        For the training of the two-dimensional Deep-TDA CV, we have used the 34 contacts described above as input of a NN with three hidden layers \{24, 12, 2\}. The target function parameters in the $\{s_1, s_2\}$ CVs space were $\mu^{tg}_A = [-5,-4.3]$, $\mu^{tg}_R = [0,4.3]$, $\mu^{tg}_M = [5,-4.3]$, to have the three states on the vertices of an equilateral triangle with side 10, and $\sigma^{tg}= [0.2, 0.2]$, giving $\Delta=35$ for each pair of states.
        The hyperparameters for the loss function (Eq.\ref{eq:loss_multi}) were $\alpha=1$ and $\beta=250$.
        In the OPES simulations we have biased the Deep-TDA CVs using \texttt{SIGMA=0.2,0.2} to match the standard deviations along the two CVs of the basins in the target distribution, \texttt{BARRIER=240} kJ/mol and \texttt{PACE=500}.
        We averaged over 4 statistically independent 10ns simulations.
        
        \begin{figure}[h!]
            \centering
            \includegraphics[width=0.4\linewidth]{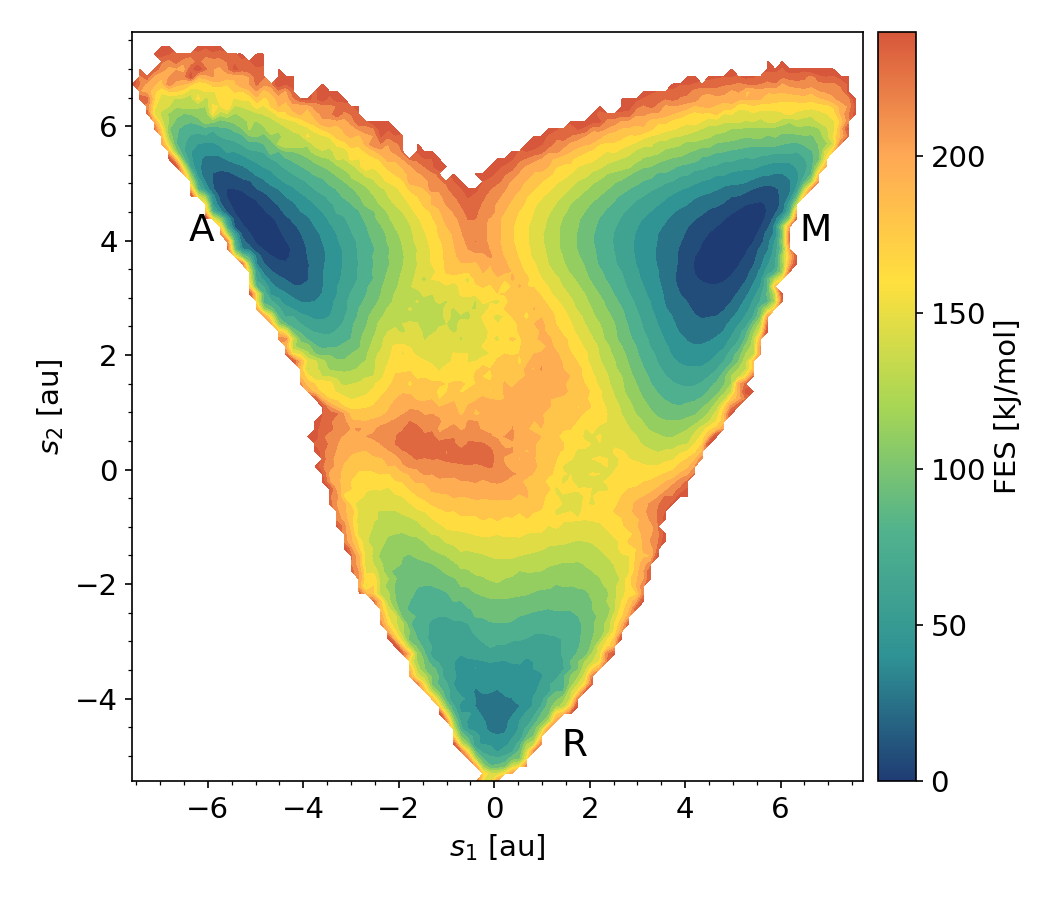}
            \includegraphics[width=0.4\linewidth]{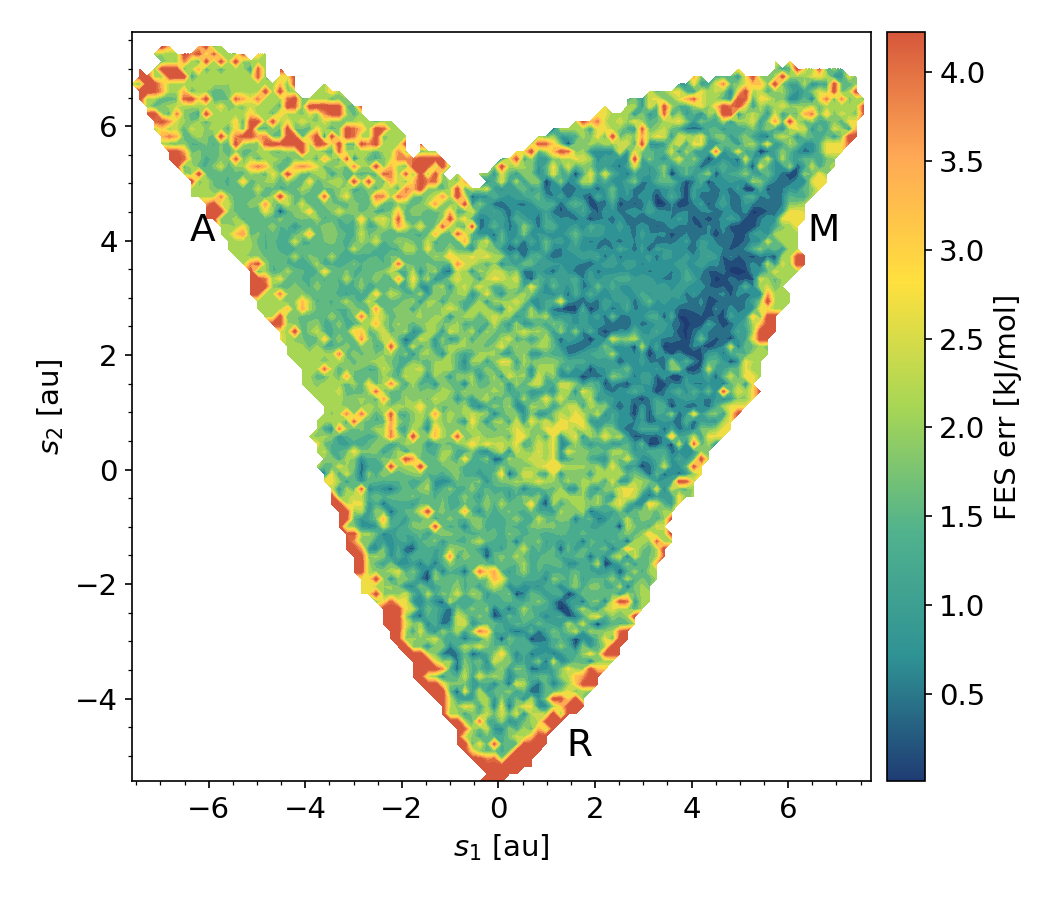}
            \caption{FES estimates for the hydrobromination of propene reaction in the plane of the Deep-TDA CVs $s_1$ and $s_2$, as reported in Fig.\ref{fig:hbro_2D} in the main text, with the associated error map. The reagents basin \textbf{R} is at the bottom, the anti-Markovnikov product \textbf{A} on the top-left position, the Markovnikov \textbf{M} one on the right.}
            \label{fig:S_hbro_std}
        \end{figure}
        
        In Fig.\ref{fig:S_hbro_std} we report the same contour plot of the FES in the Deep-TDA CVs plane given in the main text (Fig.\ref{fig:hbro_2D}) with the associated error map. As pointed out in the main text , since interconversion events between the two products are rare, the interconversion transition region is poorly sampled and the corresponding error is large.

    \paragraph{One-dimensional CV}
        For the training of the one-dimensional Deep-TDA CV, we have used the 34 contacts described above as input of a NN with three hidden layers \{24, 12, 2\}. The target function parameters were $\mu^{tg}_A = -14$, $\mu^{tg}_R = 0$, $\mu^{tg}_M = 14$ and $\sigma^{tg}=0.2$, giving $\Delta=50$ for each of the products with respect to the reagents basin.
        The hyperparameters for the loss function (Eq.\ref{eq:loss}) were $\alpha=1$ and $\beta=250$.
        In the OPES simulations we have biased the Deep-TDA CV using \texttt{SIGMA=0.2} to match the standard deviation of the basins in the target distribution, \texttt{BARRIER=240} kJ/mol and \texttt{PACE=500}.
        We averaged over 6 statistically independent 10ns simulations.
        \begin{figure}[h!]
                \centering               \includegraphics[width=0.5\linewidth]{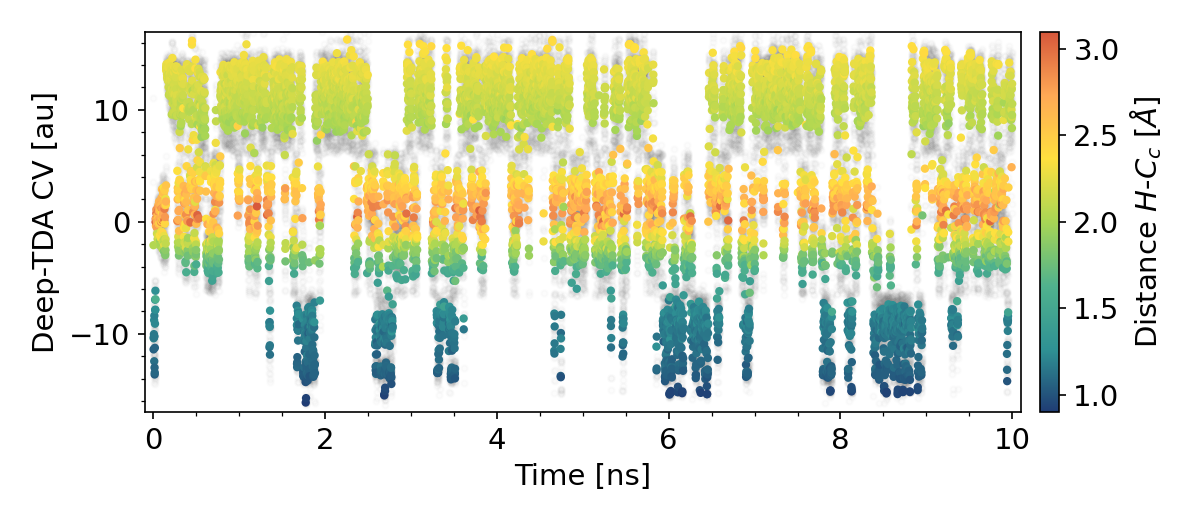}
                \caption{ 
                Running average of the Deep-TDA CV colored according to the distance between the central carbon atom $C^c$ and the reactive hydrogen $H$. The average is computed every 5 configurations. The complete time series is given in transparency in the background.}
                \label{fig:hbro_running}
            \end{figure}
        
        In fig.\ref{fig:hbro_running}, in order to highlight the three states, we report a running average of the Deep-TDA value over time colored according to the distance between the central carbon atom $C^c$ and the reactive hydrogen H.
        
\begin{figure} [h!]
            \centering
            \includegraphics[width=0.45\linewidth]{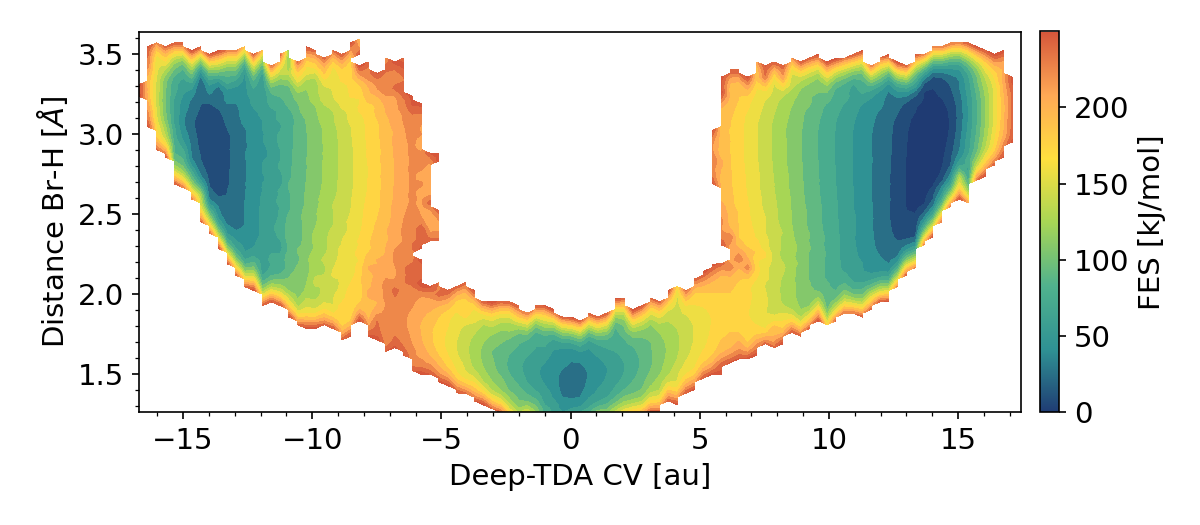}
            \includegraphics[width=0.45\linewidth]{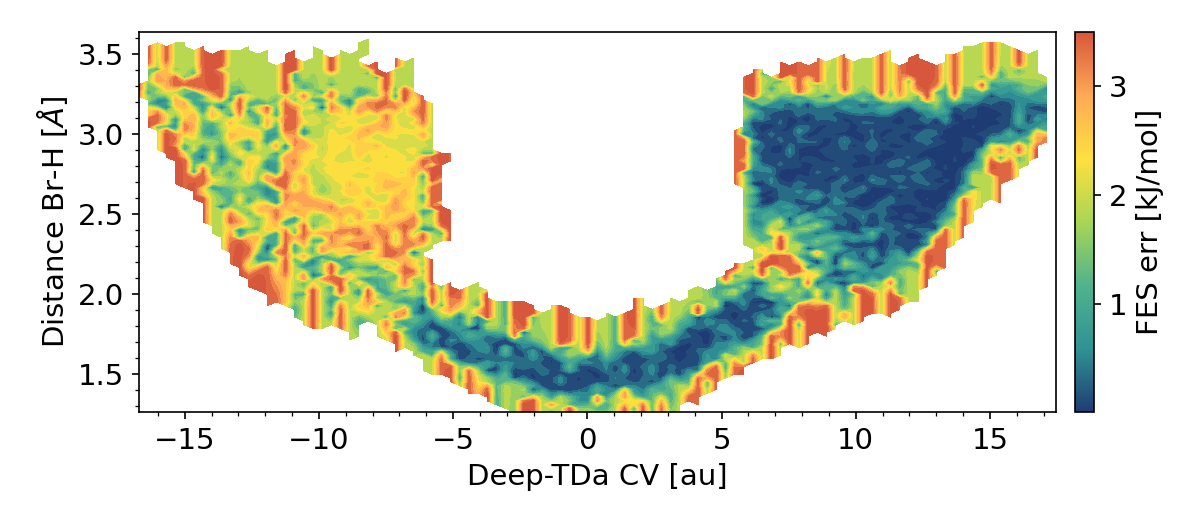}
            \caption{Contour plot of the FES estimate with the associated error map for the hydrobromination of propene in the Deep-TDA CV and H-Br distance plane. From left to right: Anti-Markovnikov product, reagents and Markovnikov product}
            \label{fig:S_hbro_2D}
        \end{figure}

        The contour plot of the FES in the Deep-TDA CV and Br-H distance plane is reported in Fig.\ref{fig:S_hbro_2D} with the associated error map, clearly showing the three basins.
        
\section{Double intramolecular proton transfer in \\ 2,5-diamino-1,4-benzoquinone}
    \vspace{-5pt}
    The intramolecular proton transfer simulations have been carried out using the CP2K-7.1\cite{kuhne2020cp2k} software package patched with PLUMED2.7 at PM6 semi-empirical level. 
    The integration step was $0.5$fs and we used the velocity rescaling thermostat set at 300K with a time constant of 100fs.
    
    The coordination numbers used for the training have been computed with PLUMED in the form 
    \begin{equation}
    s_{ij}(r) = \frac{1 - \left(\frac{r_{ij}}{\sigma_{ij}}\right)^n}{1 - \left(\frac{r_{ij}}{\sigma_{ij}}\right)^m}
    \end{equation}
    They have been computed on each heavy atom with respect to the element couples: CC, CN, CO, CH, NH, OH. 
    The $\sigma_{ij}$ parameters have been:
    $\sigma_{CC}$=1.7{\AA}, $\sigma_{CN}$=1.7{\AA}, $\sigma_{CO}$=1.6{\AA}, $\sigma_{CH}$=1.2{\AA}, $\sigma_{NH}$=1.1{\AA}, $\sigma_{OH}$=1.2{\AA}. The exponent in the same function have been n=6 and m=8 to have a smooth behavior over a wide range of distances.
    
    For the one-dimensional Deep-TDA CV training, we used these 28 coordination numbers as input for a NN with three hidden layers \{24, 12, 1\}. The target function parameters were $\mu^{tg}_R = -14$, $\mu^{tg}_I = 0$, $\mu^{tg}_P = 14$ and $\sigma^{tg}=0.2$, giving $\Delta=50$ for the R and P states with respect to the intermediate I.
    The hyperparameters for the loss function (Eq.\ref{eq:loss}) were $\alpha=1$ and $\beta=250$.
    
    In the OPES simulations we have biased the Deep-TDA CV using \texttt{SIGMA=0.2} to match the standard deviation of the basins in the target distribution, \texttt{BARRIER=120} kJ/mol and \texttt{PACE=500}.
    The average values in the main text are obtained averaging over 10 statistically independent 10ns simulations.
    
\begin{figure}[h!]
        \centering               \includegraphics[width=0.8\linewidth]{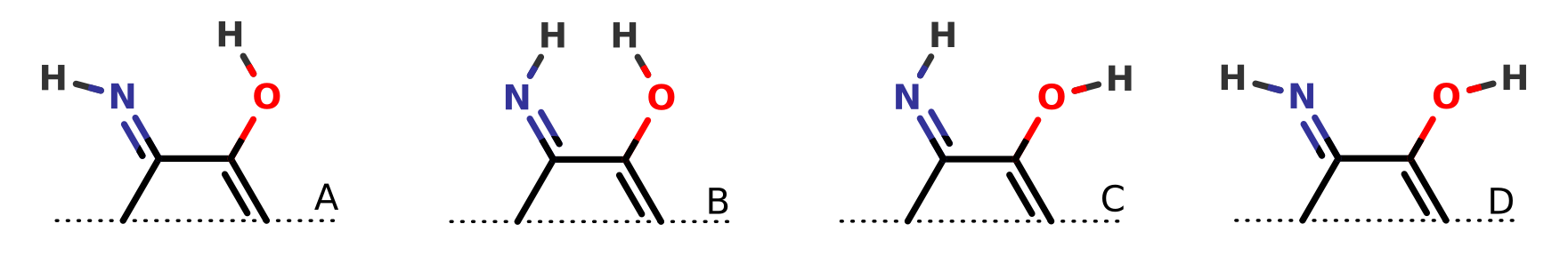}
        \caption{ 
        Sketch of the possible relative dispositions of the hydrogens in E and I metastable states. A-type conformers are the most likely.}
        \label{fig:S_benzo_conformers}
    \end{figure}

    As pointed out in the main text, different conformers can be distinguished within the I and P metastable basins. The possible relative orientation of the hydrogens of the heteroatoms are sketched in Fig.\ref{fig:S_benzo_conformers}. The most sampled conformers adopt the A-type configuration, as it is the one that favours the most the exchange of hydrogen between the two functional groups and is also stabilized by an hydrogen bond.
    
\vspace{-10pt}
\section{Code availability}
    The code used for the training of the Deep-TDA CVs is available in a tutorial Google Colab notebook \href{https://colab.research.google.com/drive/1TO7bAkmIznsdfea2i5NXfNtytJrnkkIt?usp=sharing}{Deep-TDA tutorial} \footnote{https://colab.research.google.com/drive/1TO7bAkmIznsdfea2i5NXfNtytJrnkkIt?usp=sharing}.
    
    The files with the descriptors used as input for the NN training, the input files for the simulations and the trained models to reproduce the results of this paper can be found on \href{https://github.com/EnricoTrizio/TargetedDiscriminantAnalysisCVs}{Github} \footnote{https://github.com/EnricoTrizio/TargetedDiscriminantAnalysisCVs} and in the PLUMED-NEST repository, as \href{https://www.plumed-nest.org/eggs/21/028/}{plumID:21.028}\footnote{https://www.plumed-nest.org/eggs/21/028/}.
    
\bibliography{bibliography}